# Nanometer-precision tracking of adipocyte dynamics via single lipid droplet whispering-gallery optical resonances

Rok Podlipec[1*], Ana Krišelj[1], Maja Zorc[1], Petra Matjan Štefin[2,3], Siegfried Usaar[4,5], Matjaž Humar[1,6,7]

[1] Department of Condensed Matter Physics, Jozef Stefan Institute, Jamova 39, SI-1000 Ljubljana, Slovenia

[2] Department of Biochemistry and Molecular and Structural Biology, Jozef Stefan Institute, Jamova cesta 39, SI-1000 Ljubljana, Slovenia

[3] Jozef Stefan International Postgraduate School, Jamova cesta 39, SI-1000 Ljubljana, Slovenia

[4] Research Unit Adipocytes & Metabolism (ADM), Helmholtz Diabetes Center, Helmholtz Zentrum München, Germany; Research Center for Environmental Health GmbH, Neuherberg 85764, Germany

[5] German Center for Diabetes Research (DZD), 85764, Neuherberg, Germany

[6] Faculty of Mathematics and Physics, University of Ljubljana, Jadranska 19, SI-1000 Ljubljana, Slovenia

[7] CENN Nanocenter, Jamova 39, SI-1000 Ljubljana, Slovenia

Corresponding author: rok.podlipec@ijs.si

## Abstract

Biophotonics—and more recently, biointegrated photonics—offer transformative tools for probing cellular processes with unprecedented precision. Among these, whispering gallery mode (WGM) resonators—optical microcavities formed in spherical structures—have emerged as powerful biosensors and intracellular barcodes. Lipid droplets (LDs), with their high refractive index and intrinsic spherical geometry, are ideal candidates for supporting intracellular lasing. Although lasing in LDs has been previously demonstrated, it has not yet been harnessed to study live cell biology. Here, we report the first use of WGM resonances in LDs of live primary adipocytes, employing a continuous-wave (CW) laser at powers below the biological damage threshold. By measuring these resonances, we achieved nanometer-scale precision in size estimation, enabling real-time observation of rapid LD dynamics and deformations on the minute scale—far beyond the spatio-temporal resolution of conventional microscopy. We systematically characterized this photonic sensing approach, demonstrating its ability to resolve adipocyte heterogeneity, monitor lipolytic responses to forskolin and isoproterenol, and detect early signs of cell viability loss—well before conventional assays. This proof-of-concept establishes intracellular LD WGM resonances as a robust platform for investigating live single-cell metabolism. The technique enables rapid, cost-effective assessment of adipocyte function, reveals cell-to-cell variability obscured by bulk assays, and lays the foundation for high-throughput analysis of metabolism- and obesity-related diseases at both cellular and tissue levels.

Biointegrated photonics and biophotonics are rapidly emerging fields in cellular sensing, leveraging advanced optical technologies to study and manipulate biological processes with exceptional precision [1]. One of the highly promising biosensing strategy employs whispering gallery mode (WGM) resonators [2,3], which exploit optical resonances in spherical objects, such as microspheres and microdroplets. Light waves induced and propagated within these cavities undergo continuous internal reflection along the concave surface, producing constructive interference [4]. The resonant conditions—capable of achieving extremely high quality (Q) factors—depend on the refractive index contrast between the cavity and its surrounding environment and exponentially on the microcavity size [5]. These unique properties have enabled a broad spectrum of applications in biological and physical sensing [4,6–8], offering remarkable sensitivity for detecting subtle changes at the single-cell [9] and single-molecule levels [10–12]. Applications range from detecting single proteins and silica nanobead binding [10], to plasmon-enhanced sensing with nanorods for short nucleic acid strands [11], and even single-virus tracking [13]. More recently, significant attention has been directed toward intracellular probing techniques for cell tagging, barcoding, and tracking [14–19], cavity-enhanced bioluminescence [20], and investigations into cellular (patho)physiology, including cardiac contractility [21] and molecular binding [22].

Despite the promise of this emerging research frontier, intracellular studies utilizing biointegrated microlasers remain limited, particularly in addressing biologically relevant questions such as complex cell heterogeneity in disease contexts [9]. One highly underexplored yet potentially transformative approach involves lasing on endogenous cellular structures. Lipid droplets (LDs), owing to their high sphericity and elevated refractive index relative to the surrounding cytoplasm, are capable of supporting intracellular lasing [14]. Lasing in LDs is feasible for droplets larger than approximately 25 µm, where radiative losses are sufficiently minimized. This could make adipose tissue (AT)—which contains LDs of the sizes ranging up to 100 µm and more in mature adipocytes [23]—an ideal candidate for such applications. However, to date, intracellular lasing in LDs has not been applied to biological studies of adipocytes.

LDs primarily serve as reservoirs for lipids—essential components for maintaining metabolic energy reserves and supplying lipids for cellular membranes [24]. Although central to adipocyte function, the role of LD metabolism and its dysregulation in metabolic diseases remains underexplored [25]. It is now recognized that LDs play multiple roles in systemic homeostasis and obesity-related pathologies [26–29], extending far beyond fat storage. They act as dynamic hubs for lipid management, integrating metabolic signals and lipid fluxes with diverse cellular homeostatic and stress responses [30]. Despite their critical role in AT regulation, the dynamic nature and biological activity of LDs are still poorly understood [31]. Current data are largely limited to single-cell gene expression analyses [32] and bulk functional assays, such as lipolysis kits applied to large populations of adipocytes [33]. These approaches fail to capture the heterogeneity of LD function and dynamics both within and between individual cells. Moreover, averaging LD behavior across cell populations obscures the contributions of individual organelles, limiting our understanding of their nuanced/complex roles.

To overcome these limitations novel optical methods are needed to assess LD dynamics in living, single primary adipocytes from animal or human sources. In recent years, several experimental approaches have emerged, including morphology-oriented live-cell and single-cell imaging of LDs using fluorescence microscopy [34,35], the development of advanced fluorescent probes [36], label-free Raman microscopy [37,38], as well as machine learning–based tools and analysis [39]. Although these techniques have enabled the study of metabolic processes at cellular and subcellular levels previously inaccessible [40], they remain constrained by the optical resolution limit of ~250 nm (Rayleigh criterion), which prevents detection of morphological changes below this threshold. Limited spatial resolution also restricts temporal resolution, making it difficult to capture rapid, nanometer-scale changes in LD size—processes critical for understanding early metabolic shifts and predicting long-term outcomes of LD activity. LD growth (e.g., induced lipogenesis [41]) and shrinkage (e.g., induced lipolysis [42]) are often slow, occurring over hours, days, or even weeks before substantial size changes become detectable via live-cell microscopy [43]. Achieving nanometer precision in LD size measurements would enable near-instantaneous detection of growth or shrinkage, offering valuable insights into short-term metabolic dynamics.

To address these challenges, we present and characterize, for the first time, WGM optical resonances in natural intracellular lipid droplets as a means of biological sensing in adipocytes. Building on our previously established WGM-based methodology for embedded lasing in live cells [14], this approach offers exceptionally high spatial (nanometer) and temporal (minute-scale) resolution for precise tracking and investigating individual LD size and dynamics. While our previous study focused solely on demonstrating lasing in adipocytes, here we systematically evaluate the methodological, physical and biological aspects of photonic sensing in live adipocytes, highlighting both its strengths and limitations. This includes a detailed assessment of potential laser-induced effects and the optimization of laser parameters, which identify low-power continuous-wave (CW) lasers as the most suitable source for biological applications. Rather than focusing on detailed biological interpretations, our primary aim is to establish a robust methodological foundation for future studies. These findings pave the way for advanced metabolic investigations into the highly heterogeneous and dynamic nature of adipocytes and adipose tissue.

## Materials and methods

MATERIALS.

Visceral fat (isolated from FVB/N female mice), 1x phosphate buffer saline (PBS, Gibco), Dulbecco's Modified Eagle's Medium (DMEM, Gibco), Collagenase IV (Gibco), Fatty Acid Free Bovine Serum Albumin (BSA FAF, Sigma Aldrich), Fetal Bovine Serum (FBS, Gibco), Pyrromethene 597 (Exciton Luxottica), verapamil (Spirochrome), SYTOX Deep Red (Thermo Fischer Scientific), CellMask Deep Red (Thermo Fischer Scientific), Forskolin (Sigma Aldrich), Triacsin (Cayman).

ANIMALS.

Mice were used in accordance with the Administration of the Republic of Slovenia for food safety, veterinary and plant protection (permit number: U34401-5/2022/15). Procedures for animal care and experiments were in accordance with the "Guide for the Care and Use in Laboratory Animals".

METHODS.

*Adipocyte isolation*. Visceral AT was isolated from FVB/N female mice. After the removal of larger veins, AT was cut into smaller pieces (1-2 $mm^3$) and washed three times in cold PBS. Pieces were transferred in a 25 mL centrifuge tube with DMEM, complemented with penicillin (100 IU/mL), streptomycin (100 µg/mL), 1 mg/mL Collagenase Type IV, and 1% (w/v) BSA (FAF). 2 mL of the collagenase solution was used per 1 g of fat tissue. Minced fat tissue was then incubated at 37 °C for 30-40 min and gently shaken every 10 min to allow for adequate disaggregation and isolation of individual adipocytes, preserving their viability. After digestion was complete, the suspension was mixed on a vortex mixer for 10 seconds to release the remaining cells from the tissue and then passed through a sieve with 100 µm pores. Collagenase was neutralized with the same volume of FBS. The cell suspension was then allowed to separate into layers at 37 °C for 15 minutes due to density differences. The top lipid layer and the bottom medium layer were discarded, followed by carefully pipetting of mature adipocytes into a fresh centrifuge tube, where they were washed 3 times with cell culture medium (DMEM, complemented with penicillin (100 IU/mL), streptomycin (100 µg/mL), 4mM L-Glutamine, 10 % FBS, 0.5 µg/mL insulin, 0.4 ng/mL dexamethasone). Mature adipocytes were cultured in 12-well plates at 37 °C and 5% $CO_2$. Each well in a 12-well plate was filled with 1 mL of cell culture medium and 200 μL of dense suspension of isolated adipocytes. Medium without insulin and dexamethasone was used for further experiments.

*Adipocyte labeling for optical sensing and dynamics studies*. The cell suspension was mixed with 1.5 μg/ml (4 μM) pyrromethene 597 BODIPY laser dye for labeling of LDs to enable lasing applications. After a few hours of incubation at 37°C and 5% $CO_2$, the cell media containing LDs label was exchanged with the cell media containing CellMask Deep Red (2.5 μg/ml, 2000x diluted stock solution) and SYTOX Red (5 nM, 1000x diluted stock solution) for 15 minutes incubated on 37°C and 5% CO2. The volume of 1 ml suspension was then exchanged four times with insulin-free cell media to remove non-labeled stains. For the optimal experimental conditions, ceiling translucent trans-well inserts (3 um pores, Falcon) were carefully immersed and fixated inside the 12-wells, where a significant part of the densely populated floating adipocytes remained confined beneath the porous surface (Figure 1). For studying adipocyte dynamics and lipolytic metabolic activity in response to external stimuli, a cell medium containing lipolytic agents, forskolin, and isoproterenol, was gently poured and mixed in the trans-well chamber over time. A combination of 20 μM forskolin and 5 μM triacsin was used to stimulate lipolysis [44] and prevent the regeneration of triglycerides [45]. In contrast, 10 μM isoproterenol, acting through a different mechanism of cellular signaling [46], was used to stimulate lipolysis. Additionally, 5 μM verapamil was added to forskolin solution to check possible lipolytic change by blocking calcium (influx) channels [47] important for regulation of the lipid metabolism [48,49]. For control samples, no lipolytic agents were added in the trans-well chamber. A 12-well plate with Transwell inserts was then transferred into the stage-top incubator (H301-K-FRAME, Okolab) mounted on an inverted microscope (Nikon ECLIPSE Ti2) for LD dynamics studies on live adipocytes. The experiment was run for a total of 6 to 24 hours, with lipolytic agents being added 2-3 hours after the start.

*Optical setup*. Adipocyte LDs, as natural intracellular optical microresonators, were excited at their circumference using a nanosecond (ns) pulsed (Opotek, OpoletteTM 355) and compact diode CW (Thorlabs) lasers, both set to a wavelength of $\lambda$ = 532 nm, optimized for BODIPY laser dye excitation. To achieve the optimal lasing on live adipocytes below the threshold for photodamage, the energy per pulse of the pulsed laser with a 20 Hz repetition rate was typically 0.1-0.5 μJ (with an irradiance of $I_{\mathrm{pulse}} < 10^9$ W/cm$^2$), while the power of the CW laser was 0.1 μW with an irradiance of $I_{\mathrm{pulse}} \approx 10$ W/cm$^2$. The lasers were aligned in the back port of the microscope. For optimal focusing and detection of WGM resonances, a 20× objective (*NA* = 0.45) and exposure times of 0.5 to 5 seconds were used, respectively. WGM spectra were captured using an imaging spectrometer (Shamrock SR-500i, Andor) with a diffraction grating of 1.200 lines per millimeter, resulting in a resolution of 0.07 nm. A 50 μm wide slit aperture was used at the spectrometer entrance for optimal confocality and signal-to-noise (S/N). Wide-field fluorescence imaging of cell organelles, LDs, plasma membrane, and nuclei was acquired simultaneously in separate fluorescence channels with a digital camera (sCMOS Zyla 4.2, Andor) using LED source illumination (CoolLED, pE-300 white), dichroics and band-pass filters (all Semrock).

*Image/spectral analysis*. The measured spectra of pump-induced lasing or high S/N are the superposition of (differently polarized) transverse electric (TE) and/or transverse magnetic (TM) spectral eigenmodes/WGMs [3]. Their positions were calculated according to the first order radial modes approximation description [50,51], where each peak position was fitted with the following function:

$$L(\lambda) = \sum_{i_{TM},i_{TE}} \frac{1}{2\pi} \frac{\Gamma}{\left(\lambda - \lambda_{i_{TM},i_{TE}}\right)^2 + (1/2\Gamma)^2} \tag{1}$$

$\lambda_{i_{TM},i_{TE}}$ are the spectral peak positions defined by the mode numbers (*m*), polarization, refractive index ratio, the LD size, etc., and $\Gamma$ is a parameter specifying function width. We assumed that both internal and external refractive indices are constant and directly calculated the LD size. The assumption of constant refractive indices is justified in the continuation. The examples of spectral fits to the background-subtracted and normalized raw data in a time experiment are shown in Figures S1 and S2a. In cases of low S/N, where the spectra resembled a sinusoidal shape, we developed an empirical model that nicely fits the experimental data after removing the background signal. Such broadened spectra are typical for a large number of measured adipocytes with non-complete sphericity and smoothness of the LD surface. The corresponding empirical model is:

$$f(\lambda) = A\sin[(k(1 - B(\lambda - \lambda_{min})^2)\lambda + \phi] \tag{2}$$

where *A* is the intensity of the WGM peaks, *k* is the wavenumber of the quasi-sinusoidal function at the left boundary of the measured spectral interval at $\lambda_{min}$ and $\phi$ is the phase shift required to properly align the fitting function to the measured WGM resonances. *B* is the scaling constant to properly fit the nonlinear (quadratic) dependence of the spacing between consecutive resonant modes, the so-called free spectral range (FRS), with wavelength *λ. B* can be used to extract the refractive index dispersion with the wavelength $\Delta n(\lambda)$. Due to typically smaller $\Delta n(\lambda)$ than the measured peak position uncertainty in low S/N spectra obtained in multiple cases, we simplified the model. Since the WGM resonances are periodic in inverse wavelength, the model can be adapted as follows:

$$f(\lambda) = A\sin[k'\lambda^{-1}] \tag{3}$$

where *k*' is a scaling constant that encodes the proportionality with FSR = $1/\pi n_{\mathrm{eff}} d$, where $n_{\mathrm{eff}}$ is the effective refractive index of LD potentially influenced by the surrounding medium through the evanescent field and $d$ is the LD diameter. From the peak positions of the fitted model (example in Figure S2b), we calculated FSR between individual peaks across the spectra ($\lambda_{m+i}$) to determine $d$ and, consequently, the change in LD diameter ($\Delta d$) over time using the equation:

$$\frac{\sum_{i=1 \dots N}\left(\frac{1}{\lambda_{m+i}} - \frac{1}{\lambda_{m+i-1}}\right)}{N} = \frac{1}{\pi n_{eff} d} \tag{4}$$

The accuracy of the calculated $\Delta d$ was validated optically (Figure S3), showing excellent correlation with our model under the assumption of a constant effective refractive index ($n_{\text{eff}}$) throughout the experiment (Figure 2). This confirmed that $\Delta d$ can be precisely quantified well below the optical resolution limit using WGM fitting alone, whereas direct optical measurements are sensitive only to $\Delta d$ above ~1 μm and are accompanied by large errors.

## Results and discussion

### Experimental approach for nanometer-precision optical sensing of mature adipocytes

Adipocytes isolated from mouse or human adipose tissue represent one of the most challenging biological systems for advanced microscopy techniques. Their study requires careful optimization of both experimental conditions and optical sensing to capture fast, dynamic processes. The unique properties of adipocytes—particularly their highly spherical shape and low density—contribute to their buoyancy in the cellular medium, necessitating a customized imaging platform, as previously demonstrated [52]. On one hand, their near-perfect sphericity demands three-dimensional fluorescence imaging for accurate characterization of cell morphology and, critically, cell viability. On the other hand, their buoyancy and surface confinement require an optimized experimental setup, achieved here using a translucent ceiling trans-well system (schematic in Figure 1a). This workflow enabled controlled administration of chemical reagents to stimulate metabolic activity while maintaining the immobility of non-adherent, mature adipocytes—essential for dynamic single-cell studies. We began with large-field-of-view (FoV) bright-field (BF) imaging to assess the quality of adipocyte isolation, evaluate ceiling coverage under the trans-well, and determine cell size distribution (Figures 1 and 2a). Evenly distributed pores, averaging 8 μm in diameter and visible in fluorescence imaging (see arrows), facilitated solvent diffusion from the top to the bottom chamber, achieving complete mixing and concentration equalization within one hour (Figure S4). Finally, fluorescently labeled LDs were illuminated at their circumference with a laser to excite WGMs (Figure 1c).

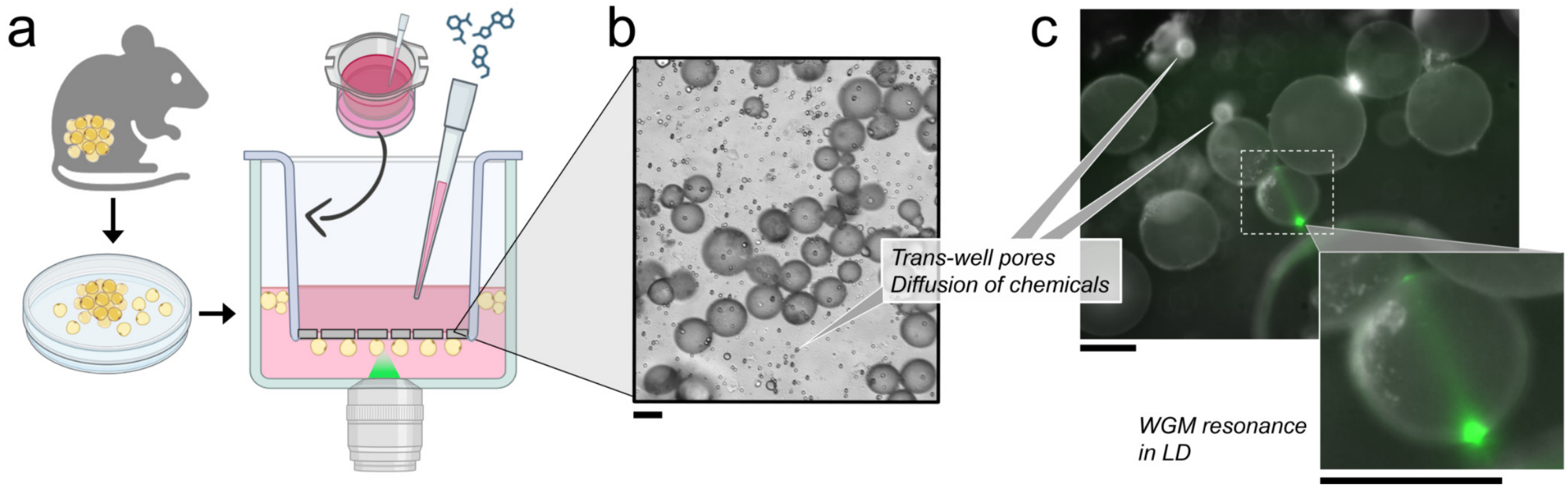

**Figure 1.** Trans-well-based experimental setup with the typical size, surface coverage, and localized optical resonance in mature adipocytes isolated from mice. (a) Schematics of a workflow and a translucent trans-well insert placed in a 12-well plate where the cells are homogeneously distributed under the surface. The plate was mounted on a stage top incubator fitted on an inverted microscope. (b) A typical example of a bright field (BF) image, showing cell distribution and size with visible 8 µm pores that enable diffusion and exchange of chemicals between the upper and lower chambers. (c) An example of a fluorescence image of adipocytes with the locally focused laser beam (in green) exciting WGM resonances in a vertical plane visible as a line and a smaller bright spot on the other side of the LD. Scale bar is 50 µm.

Prior to measuring LD size via optical resonances, we assessed adipocyte viability using plasma membrane staining (CellMask Deep Red) and nucleic acid staining (SYTOX Deep Red), enabling both direct and indirect detection of compromised membranes (Figure 2b). Variability in sample preservation across biological replicates—likely due to differences in AT from individual mice—resulted in adipocyte viability ranging from approximately 50% to 80% (Figure S5). Fluorescence staining also allowed us to evaluate the functional integrity and structural changes of adipocytes. The results presented in Figure S6 suggest potential cellular mechanisms involving plasma membrane remodeling, discussed in greater detail in Supplementary Note A. Given the complexity of the biological system under study, we focused exclusively on mature, non-differentiated adipocytes that maintained structural and functional integrity of both the plasma membrane and LDs throughout the duration of the experiment.

### Biological assessment and evaluation of LD lasing

By precisely positioning the pump laser at the perimeter of LDs, we excited the gain medium composed of Bodipy-based Pyromethene 597 (Figure 2c). The resulting fluorescent light was confined along the LD boundary via total internal reflection, as illustrated schematically in Figure 2d, producing distinct spectral features known as WGMs (Figure 2e and Supplementary Video 1). Depending on the excitation source—pulsed or continuous-wave (CW) laser—different gain conditions were achieved, resulting in either true lasing or cavity-modified fluorescence [5]. In both cases, the spectra exhibited sharp lines corresponding to WGM resonances; however, under lasing conditions, these lines were narrower and more pronounced (Figure 2e, right panel). Surpassing the lasing threshold required high concentrations of the lipophilic fluorescent dye and particularly high-energy ns pulsed laser excitation, reaching up to µJ per pulse. These settings were found to induce local photodamage in adipocytes via photoablation, as detailed in reference [53]. High-Q WGMs below the damage threshold were only achievable in adipocytes with structurally compromised LDs that exhibited near-perfect sphericity. This geometry, along with a smooth and polished LD surface, is known to support sharp resonance peaks [54] and enable sub-nanometer resolution [14]. In practice, however, the shape and surface smoothness of LDs are influenced by mechanical and tension forces exerted by the surrounding cytoskeleton—primarily a diffuse cortical actin network [55] and intermediate/vimentin filaments [56]. These forces can locally disrupt LD sphericity and surface integrity, significantly reducing the achievable Q-factor.

Only a small number of adipocytes emitted measurable WGMs when excited by a ns pulsed laser at a pulse energy of $E_{\text{pulse}} \approx 500$ nJ and a peak power density of $I_{\text{pulse}} \approx 10^9$ W/cm$^2$. These levels are approximately ten times higher than those used in our previous study [14], approaching the photoablation damage threshold [53]. Surprisingly, a cost-effective, low-power CW diode laser outperformed the pulsed laser in both spectral sensitivity and safety, as demonstrated on the same adipocyte (Figure S7). We identified the CW diode laser as the most optimal light source suitable for 100% safe WGM-based biosensing in native, mature adipocytes. Its operating parameters—carefully set just below the threshold for photochemical or photothermal effects (Figure S7)—consistently produced detectable WGMs (spectra in blue). The SNR varied depending on the concentration of the gain medium and local deviations in sphericity and surface smoothness at the LD boundary of individual adipocytes (Figure 2e, left panel; Figure S2).

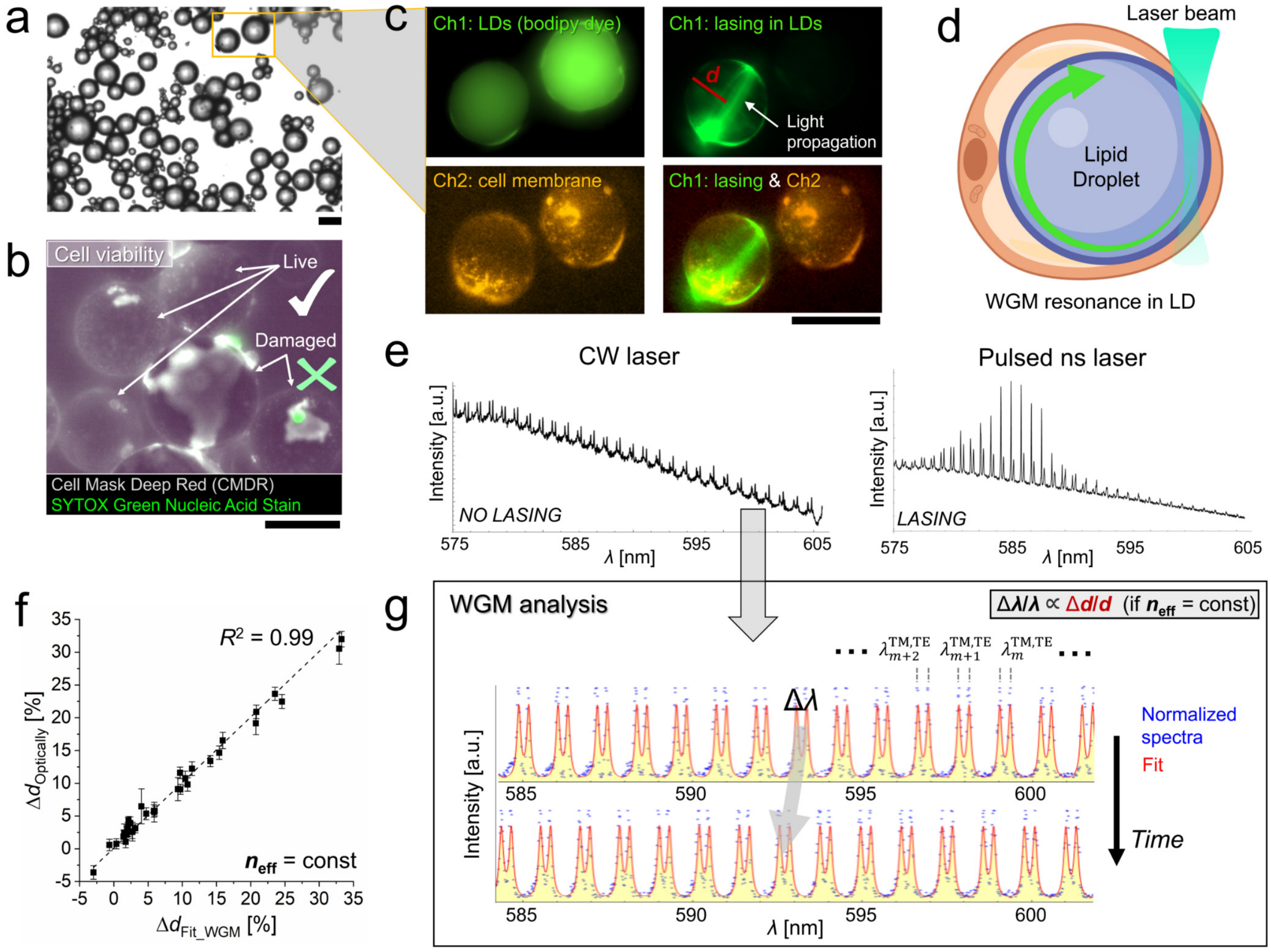

**Figure 2.** The workflow of ultraprecise adipocyte dynamics study using WGM sensing and analysis. (a) Bright field (BF) image of isolated adipocytes. (b) An example of cell viability measurement with double staining used to evaluate both structural and functional integrity. Typically, damaged cells with the labeled nuclei (in green) are accompanied by ruptured cell membranes (more images in Figure S5b and the Supplementary Video 2). (c) Upper row: fluorescence of LDs and CW laser-induced light propagation at the LD circumference observed in the same channel (Ch1, in green). Bottom row: fluorescence of cell plasma membrane observed in other channel (Ch2, in orange) and the overlay of Ch1 - lasing and Ch2. (d) Schematics of the laser excitation of the dye within the LD and the light circulation at the LD circumference forming WGM resonances. (e) Typical WGM spectra formed inside adipocyte LDs induced by CW laser (on the left) and pulsed ns laser (on the right). (f) Validation of the WGM fitting using optical measurements of the LD Δ*d* taken after one day showing a perfect correlation ($R^2$ = 0.99), where the $n_{eff}$ was kept constant over time. (g) The following analysis of WGM resonance spectra (points in blue) with typical transverse magnetic (TM) and transverse electric (TE) eigenmodes, to quantify the position of spectral peaks ($\lambda_m$) and hence the LD size (*d*) by using proper fitting (in red). Scale bar is 50 µm.

We acquired and analyzed WGM resonances induced by a CW laser. Spectra with higher SNR, exhibiting clearly distinguishable TM and TE spectral eigenmodes (Figure 2e - left panel, Figure 2g and Figure S2a), were fitted using the first-order radial mode approximation [50]. In contrast, spectra with lower SNR and significant spectral broadening (Figure S2b) were fitted using an empirical model, as detailed in the Materials and Methods section. Due to the lower Q-factor of the detected eigenmodes in both cases—compared to the high-Q WGM lasing spectra typically observed in embedded microspherical resonators—we could not achieve the same level of precision in the peak uncertainty ($\sigma_{\lambda_{i_{TM},i_{TE}}}$) and hence in the microresonator size measurement previously reported [14]. By applying spectral fitting to eigenmodes with profiles resembling Gaussian shapes, and using a well-defined model for peak position error (dependent on spectrometer resolution, sampling density, and SNR) [57] we calculated the $\sigma_{\lambda_{i_{TM},i_{TE}}}$ [58], and thereby the resolution in peak position. For the spectra shown in Figure 2g and Figure S2a—with an SNR of approximately 40 and spectral width $w \approx 0.15$ nm—the WGM peak uncertainty was calculated to be $\sigma_{\lambda_{i_{TM},i_{TE}}} \approx 0.005$ nm, while for the spectra with lower SNR of approximately 10 shown in Figure S2b, it was calculated $\sigma_{\lambda_{i_{TM},i_{TE}}} \approx 0.027$ nm. Detailed calculation of the uncertainties is presented in Supplementary Note B.

The positions of WGM resonances are sensitive to changes in both size (*d*) and the effective refractive index ($n_{\text{eff}}$), which is the refractive index of the LD ($n_{\text{LD}}$) slightly modified by the refractive index of the nearby cell cytoplasm ($n_{\text{cell}}$). $n_{\text{cell}}$ can vary over time and has a typical spread across a population of cells of up to $\Delta n_{\text{cell}} \approx 0.0006$ [59]. Because the variation of $n_{\text{cell}}$ within a single cell is typically smaller than that observed across a population of physically and metabolically heterogeneous cells, and because WGM resonances shifts are considerably less sensitive to changes in $n_{\text{cell}}$ than in $n_{\text{LD}}$, the potential $\Delta n_{\text{cell}}$ could not be detected within the spectral peak uncertainty and the corresponding LD size resolution described in this section. If $\Delta n_{\text{cell}}$ is overestimated to be 0.001, the measured LD size change would be $\Delta d \approx 2.2$ nm for a typical 60 µm-sized LD. This is smaller than the achievable resolution. Furthermore, WGMs in live cells cannot sense changes in extracellular environment due to cytoplasmic thickness of the order of a micron. WGMs are sensitive to changes only within ~100 nm of the lipid droplet surface, corresponding to the typical penetration depth of the resonator's evanescent field and the sensitivity at this distance is already markedly reduced compared with the surface. Similar argument applies to $n_{\text{LD}}$, where molecular exchange and hence changes in chemical composition in LD are far too small to produce detectable variation in $n_{\text{LD}}$. This conclusion is further supported by optical validation of $\Delta d$ performed on a small fraction of adipocytes with optically measurable $\Delta d$ (only after prolonged exposure to lipolytic agent) (Figure 2f). The results showed excellent correlation between both methods, with $n_{\text{LD}}$ ($n_{\text{eff}}$) kept constant over time in the WGM fitting and analysis. The validation not only confirms $\Delta d$ as the primary cause of the observed WGM spectral shifts, but also demonstrates the ability to calculate $\Delta d$ with precision well below the optical resolution limit using WGM fitting alone.

In addition to $\Delta d$, spectral shifts may also arise from local deviations of the lipid droplet (LD) from perfect sphericity, potentially coupled with cellular rotation between consecutive time points. To assess such deviations, we analyzed LD size at various points along its circumference by altering the laser position (Figure S8). The observed dispersion in LD size, $\Delta d \approx 10$ nm, measured across multiple cells, was smaller than the typical spectral changes recorded during time-lapse measurements following lipolytic agent administration (Figure 3). Therefore, morphological heterogeneity is expected to have minimal impact on the accuracy of quantifying biological heterogeneity and LD dynamics over hour-scale observations. Nonetheless, it is essential to carefully account for these factors.

Mechanical drift, minor cell movement, and environmental fluctuations were evaluated as potential sources of error and found to have no significant impact on accuracy. Positional shifts during long-term experiments remained well below the imaging field of view and were routinely corrected. Because each individual measurement lasted only a few seconds under stable incubator conditions, these factors were negligible within this time frame and did not affect the recorded data.

To estimate the achievable repeatability in the measured *d*, it is necessary to account for potential scatter and noise in the multiple spectra acquired from a stable LD during a typical time-lapse experiment. WGM spectra recorded over a 3-hour interval with 0.5-hour time steps on a stable LD revealed minimal peak scatter, approximately $w_s = 0.04$ nm (Figure S9). Since the observed scatter $w_s$ slightly exceeds the WGM peak uncertainties from spectral fitting, $\sigma_{\lambda_{i_{TM},i_{TE}}}$, we used $w_s$ to estimate the achievable uncertainty of the measured $\Delta d$. Applying the relation $(\Delta d)_{min} = d\, w_s/\lambda$, we obtained an uncertainty—and thus a resolution—of approximately 3.5 nm, which is two orders of magnitude better than what is achievable with confocal microscopy. In our experimental setup, the maximum applicable numerical aperture ($NA = 0.8$) provided an optical resolution of approximately 0.5 µm. However, such measurements are accompanied by a large relative error (Figure S3), particularly for submicron- to micron-scale changes in $\Delta d$, which have been shown to dominate LD dynamics (Figure 3). Consequently, optical sensing of lipid droplet size dynamics with high precision is practically impossible.

**Adipocyte heterogeneity and rapid metabolic response revealed by the rate of LD size change**

Having demonstrated the ability to accurately quantify LD size, we now apply this capability to investigate the temporal dynamics of adipocytes. Specifically, we monitor LD size over time and evaluate its response to lipolytic stimulation with agents such as forskolin and isoproterenol.

The initial time points reveal heterogeneous behavior among individual mature adipocytes even prior to stimulation (Figure 3). Some cells exhibit lipogenesis (an increase in LD size), others undergo lipolysis (a decrease in LD size), while the remainder maintain a steady-state (homeostatic) condition. The variability in our results is consistent with recent studies employing single-cell transcriptomic profiling [32,60]. These studies demonstrate high heterogeneity in adipocyte tissue, revealing diverse cell subtypes associated with distinct physiological states. Some subtypes exhibit high lipogenic capacity, whereas others are linked to lipolysis [60].

Capturing multiple baseline time points before administering the lipolytic agent was essential for establishing a reference, enabling reliable quantification of LD kinetics following stimulation. The sparse and variable response to external stimuli reflects diverse and uneven activation across the cell population. Analysis of LD size dynamics/change ($\Delta d$) in stimulated adipocytes (Figure 3, left graphs) shows a slightly broader distribution skewed toward lipolysis. Given the inherent heterogeneity observed even in control samples, additional measurements are needed to assess the statistical significance of differences in LD size and, by extension, metabolic activity between exposed and non-exposed cells. Nevertheless, in nearly all biological replicates—each comprising approximately 10 adipocytes—we observed individual cases within the stimulated group showing substantial and rapid $\Delta d$ across consecutive time points (Figure 3, asterisks; Figure S10, arrows). An increased rate of LD size change ($\Delta d/t$) was observed within a ~30-minute interval (Figure 3, right graphs).

Stimulation with isoproterenol at different sampling intervals (Figures 3a and b) resulted in a statistically significant increase in both the population mean and the variance of the $\Delta d/t$. In contrast, stimulation with forskolin (Figure 3c) did not yield a significant difference in the population mean, but did reveal a significant increase in the variance of $\Delta d/t$. In treated cells, $\Delta d/t$ reached up to approximately 1 μm/h, compared to ~0.5 μm/h in control cells. From this, we can estimate the molar flux of molecules involved in lipolysis and lipogenesis. For a 60 μm-sized adipocyte, the average molar flux—based on observed size changes—was calculated to be approximately $4*10^{-7}$ mol/m$^2$/s. This corresponds to the transport of ~$10^9$ molecules/s across the LD surface, assuming an average triglyceride (TG) molecular volume of 2 nm$^3$ [61]. To contextualize these results, the estimated molar flux aligns to that reported in a recent study using a conventional lipolytic calorimetric assay on primary adipocytes [62]. In that study, a lipolytic rate of 160 nmol/well/h corresponds to a molar flux of $1.3\times10^{-7}$ mol/m$^2$/s, assuming that the total LD surface area in confluent cells is approximated by the surface area of the assay well. Our findings provide a quantitative assessment of the rate and extent of transient metabolic responses to lipolytic agents, revealing a complex feedback mechanism that regulates both lipolysis and LD integrity.

By comparing the lipolytic activity of isolated mature adipocytes from our study with that of model adipocytes differentiated from 3T3-L1 cells [35,39,63] under identical external stimuli, we observed a slightly weaker response in LD dynamics. Several hours of exposure to isoproterenol induced size changes of up to 2 µm (Figures 3a and 3b), corresponding to a functional readout of metabolic/lipolytic efficiency ranging from an average of 0.002 to 0.007 $\mu m^3$/min per $\mu m^2$ surface area, assuming a typical LD diameter of 60 µm. In contrast, the lipolytic efficiency in model adipocytes exposed to isoproterenol was up to ten-fold higher (average 0.02 $\mu m^3/min/\mu m^2$) [63]. Exposure to forskolin induced less pronounced size changes compared with the control (Figure 3c), up to ~0.5 µm, translating into a lipolytic efficiency of ~0.001 $\mu m^3/min/\mu m^2$. This was several-fold lower than in model adipocytes exposed to forskolin (0.004 $\mu m^3/min/\mu m^2$, assuming a typical LD diameter of 5 µm) [35,39]. The higher lipolytic efficiency in model adipocytes can be attributed to their greater responsiveness to lipolytic stimuli compared with isolated mature adipocytes, which may be metabolically compromised and possess a lower surface-to-volume ratio that limits signal transduction [63]. Nevertheless, high $\Delta d/t$ values in individual adipocytes measured within a 30-min window indicate short-period bursts of lipolytic flux after stimulation with both agents (Figure 3, asterisks), comparable to those observed in model adipocytes. These results are consistent with current understanding of LD dynamics in model adipocytes and further reveal the complexity and heterogeneity of primary adipocytes, which are more directly translatable to in vivo biology and disease contexts.

In Figure 3d, we present another example of sensing stimulated adipocyte metabolism using a combination of drugs. The sapling interval was rather long, resulting in a different metric for $\Delta d/t$ (see the asterisk), which is not comparable with the rest. However, forskolin and verapamil (in green) appear to have a greater effect on lipolysis and its rate than forskolin alone (in blue). Verapamil may enhance forskolin-induced lipolysis indirectly by blocking calcium ($Ca^{2+}$) influx channels [64]. Reduced intracellular $Ca^{2+}$ levels help sustain cAMP activity, which is essential for lipolysis. Conversely, elevated $Ca^{2+}$ concentrations are known to suppress cAMP signaling [65] and consequently inhibit lipolysis in adipocytes [66]. Further biological replicates are needed to confirm the role of verapamil in this process.

Our new approach, employing non-phototoxic CW laser, provides a valuable enhancement to existing lipolytic assays by offering a faster, more precise, and cost-effective method for sensing adipocyte responses to lipolytic agents. A key advantage is its ability to quantify responses at the single-cell level, revealing biological heterogeneity and potential intercellular interactions that conventional techniques may overlook.

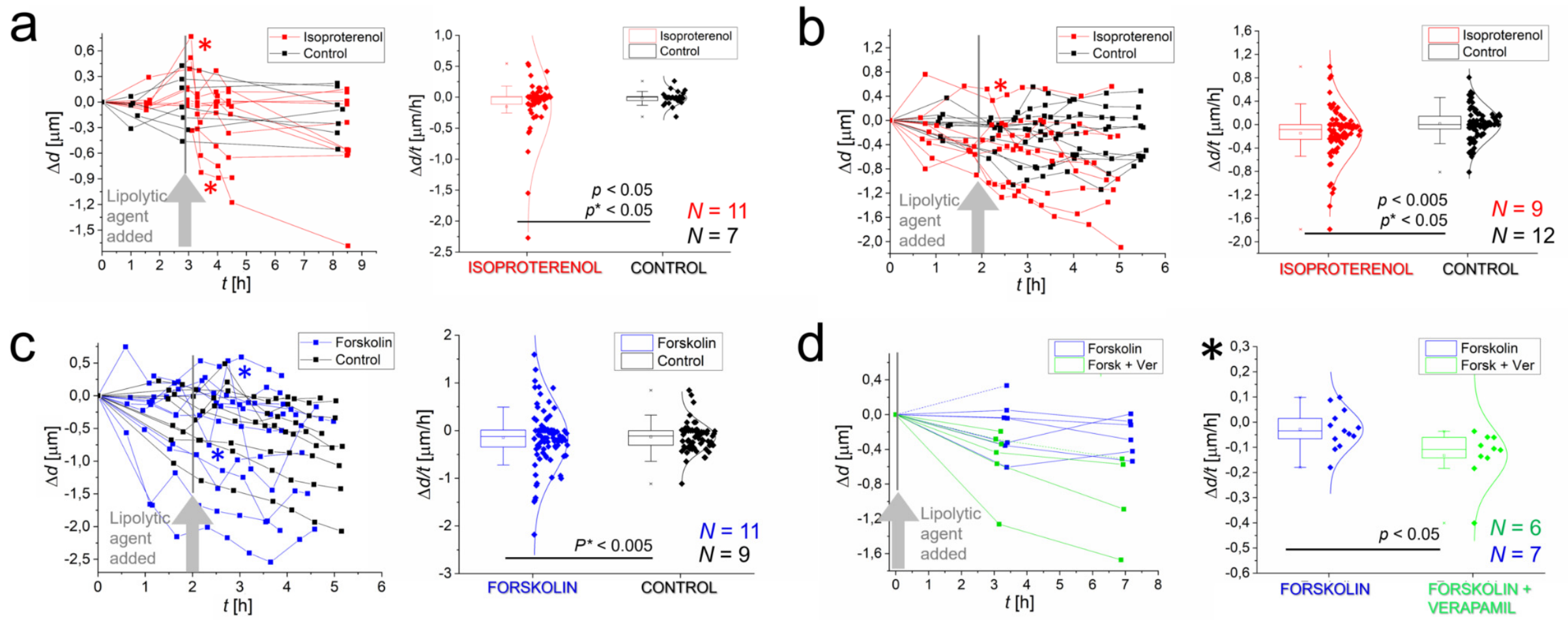

**Figure 3.** Measurements of LDs size dynamics to external stimuli introduced by isoproterenol and forskolin. (a) Isoproterenol-induced LDs size change (Δ*d*) and LDs rate of size change (Δ*d*/*t*) between consecutive time points performed on multiple cells over a longer time with a longer sampling interval. The results show significantly different population variances and suggestively different population means of the Δ*d*/*t* between the exposed cells and the control ($p = 0.05$ - two-sample t-test for null hypothesis that means are not the same; $p^* < 0.05$ - Levene's test for population variance). (b) Second biological replicate of isoproterenol-induced LDs Δ*d* and Δ*d*/*t* performed on multiple cells with a shorter sampling interval over time. The results show significantly different population variances and means of the Δ*d*/*t* between the exposed cells and the control ($p^* < 0.05$; $p < 0.05$). (c) Results of forskolin-induced LDs Δ*d* and Δ*d*/*t* performed on multiple cells in the time experiment. The results show significantly different population variance of the Δ*d*/*t* between the exposed cells and the control ($p^* < 0.005$). (d) Results of verapamil-boosted LDs Δ*d* and Δ*d*/*t* performed on multiple cells in the time experiment with long sampling interval. The results show significantly different population means of the Δ*d*/*t* between the cells exposed with one and two lipolytic agents ($p < 0.05$).

## WGM resonances enable fast diagnostics of cell viability

Throughout the study, fluorescence imaging revealed alterations in adipocyte plasma membrane morphology and integrity at the single-cell level (Figure 2b and S5). Interestingly, we noticed that these changes correlate with changes in the WGM resonances. A more detailed analysis (Figure 4) demonstrated that spectral changes occurred significantly earlier than those detectable by visual inspection or the SYTOX Deep Red viability assay. While the CW laser was optimal for measuring metabolic activity in live adipocytes, we employed a pulsed laser in this context because of its superior spectral sensitivity in distinguishing live from damaged or dead cells. This includes the sensitivity to the change in both spectral intensity and shape, which are less pronounced with a CW laser (Figure S11). To minimize the risk of photoablation damage from a pulsed laser during measurements, we used lower doses and irradiances per pulse ($E_{pulse}$ < 500 nJ; $I_{pulse}$ < $10^9$ W/cm$^2$). These conditions were generally insufficient to generate WGM resonances in live adipocytes, as shown in Figure S7.

As highlighted, WGM spectral analysis detected cellular damage prior to visual confirmation via fluorescence imaging—either directly through plasma membrane integrity (Figure 4a) or indirectly via nuclear staining (Figure 4b). Characteristic spectral transitions were observed as WGM resonances shifted from a non-lasing (sub-threshold) to a lasing mode (Figures 4a and b, blue to green spectra, indicated by the colored arrows). The full time-lapse sequence of adipocyte membrane rupture and partial LD release into the extracellular space for the example presented in Figure 4b is available in Supplementary Video 2.

WGM measurements are highly sensitive to ultrasmall changes in lipid droplet (LD) morphology, their physical properties, and the characteristics of the surrounding microenvironment. These measurements likely detect cytoskeletal remodeling—particularly actin and microtubule disruption—leading to LD relaxation toward a more spherical and smoother surface morphology, a process that precedes plasma membrane damage and subsequent cell apoptosis. This underscores the importance of mechanical signals, alongside biological markers, in assessing cell viability, as recently demonstrated [67]. A detailed analysis of the spectra shown in Figure 4c and presented in Figure S12 revealed a slight drift between the second and third time points, accompanied by a measurable increase in transverse electric (TE) and transverse magnetic (TM) mode splitting. This shift suggests a reduction in the refractive index of the LD's immediate surroundings ($n_{cell}$), primarily composed of cytoplasm. Spectral fitting indicated a decrease in $\Delta n_{cell}$ of approximately 0.028 ± 0.002, from $n_{cell}$ = 1.367 to $n_{cell}$ = 1.339, assuming $n_{LD}$ = 1.47 [68]. These findings align with both spectral and visual evidence of compromised plasma membrane integrity, where cytoplasmic dilution due to extracellular fluid influx lowers $n_{cell}$. The increased mode splitting is likely attributable to the TM mode's heightened sensitivity to perturbations in the near-field environment, owing to its deeper penetration into the evanescent field at the resonator surface, as described in prior experimental and theoretical studies [69,70]. Another example of this phenomenon is shown in Figure S13, where mode splitting correlates strongly with plasma membrane rupture. In this case, the refractive index of the cytoplasm decreased by $\Delta n$ = 0.032 ± 0.002, from approximately $n$ = 1.392 to $n$ = 1.36, indicating significant hydration and cytoplasmic loss. Through precise WGM spectral quantification in adipocytes, this study not only captures rapid metabolic dynamics but also reveals the physiological state of cells prior to conventional viability assays.

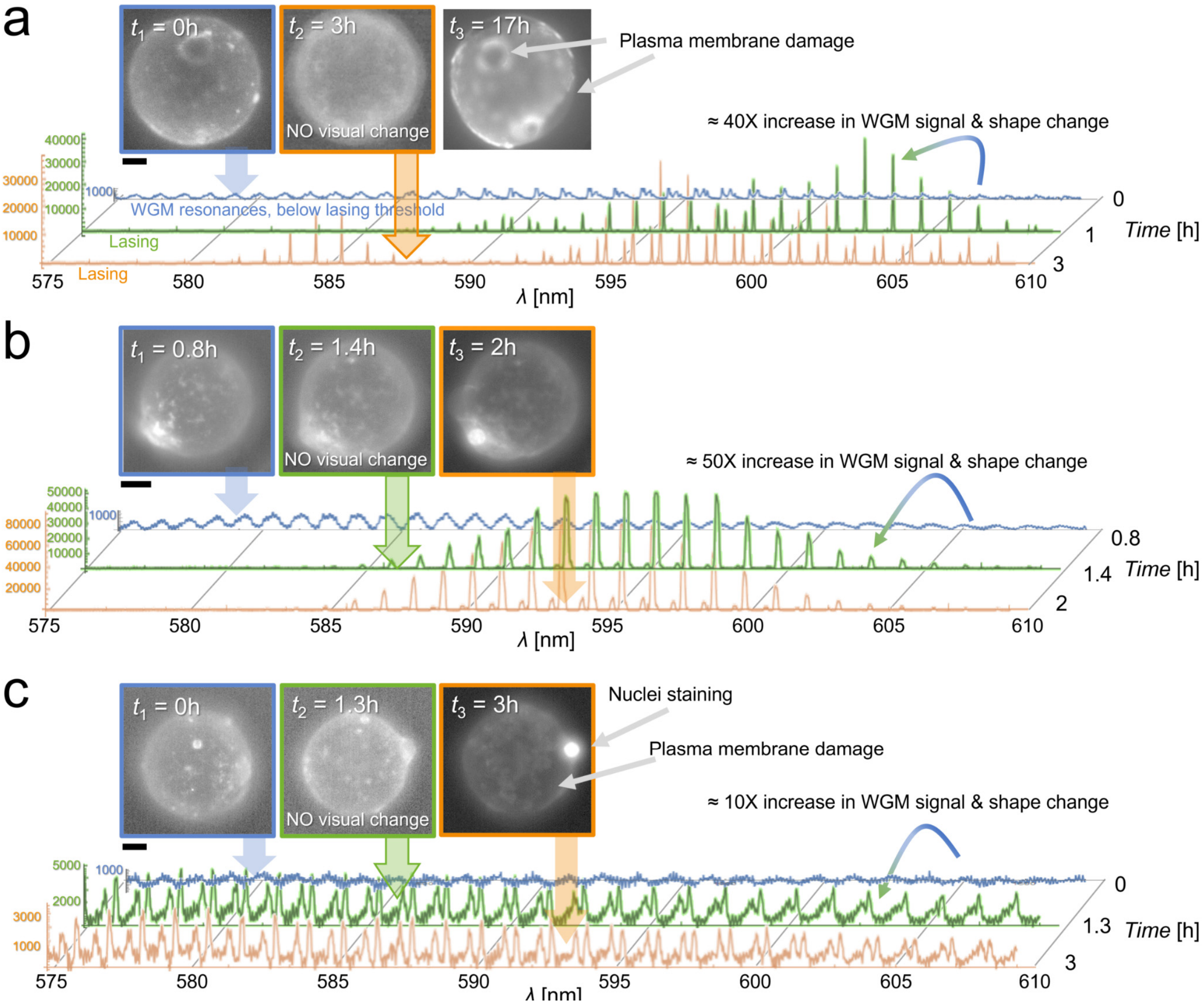

**Figure 4.** WGM spectra-based rapid diagnostics of adipocyte viability. (a) Fluorescence images of the same cell at three time points, stained with plasma membrane CellMask Deep Red dye and the corresponding changing WGM spectra color-coded in blue, green and orange. (b-c) Two more examples on adipocytes which undergo plasma membrane rupture observed additionally with the SYTOX Deep Red nuclei staining dye as a viability assay. Again, spectral change observed between first and second time points was detected before any morphological change (image outlined in green) and before being diagnosed with the viability assay (image outlined in orange). The time-lapse of the adipocyte damage, including the excretion of LD content, is provided in the Supplementary Video 2. The scale bar is 10 μm.

Time-lapse experiments across multiple biological replicates also revealed that damaged cells—initially identified by spectral changes and later confirmed by plasma membrane disruption (Figure 4)—exhibited a loss of LD dynamics, maintaining a constant LD size over time (Figure 5). The distinction between damaged and viable adipocytes is clearly demonstrated by comparing the distributions of LD size dynamics (black vs. orange). These findings reinforce that LD dynamics, monitored over several hours, serve as a reliable indicator of cellular state and viability.

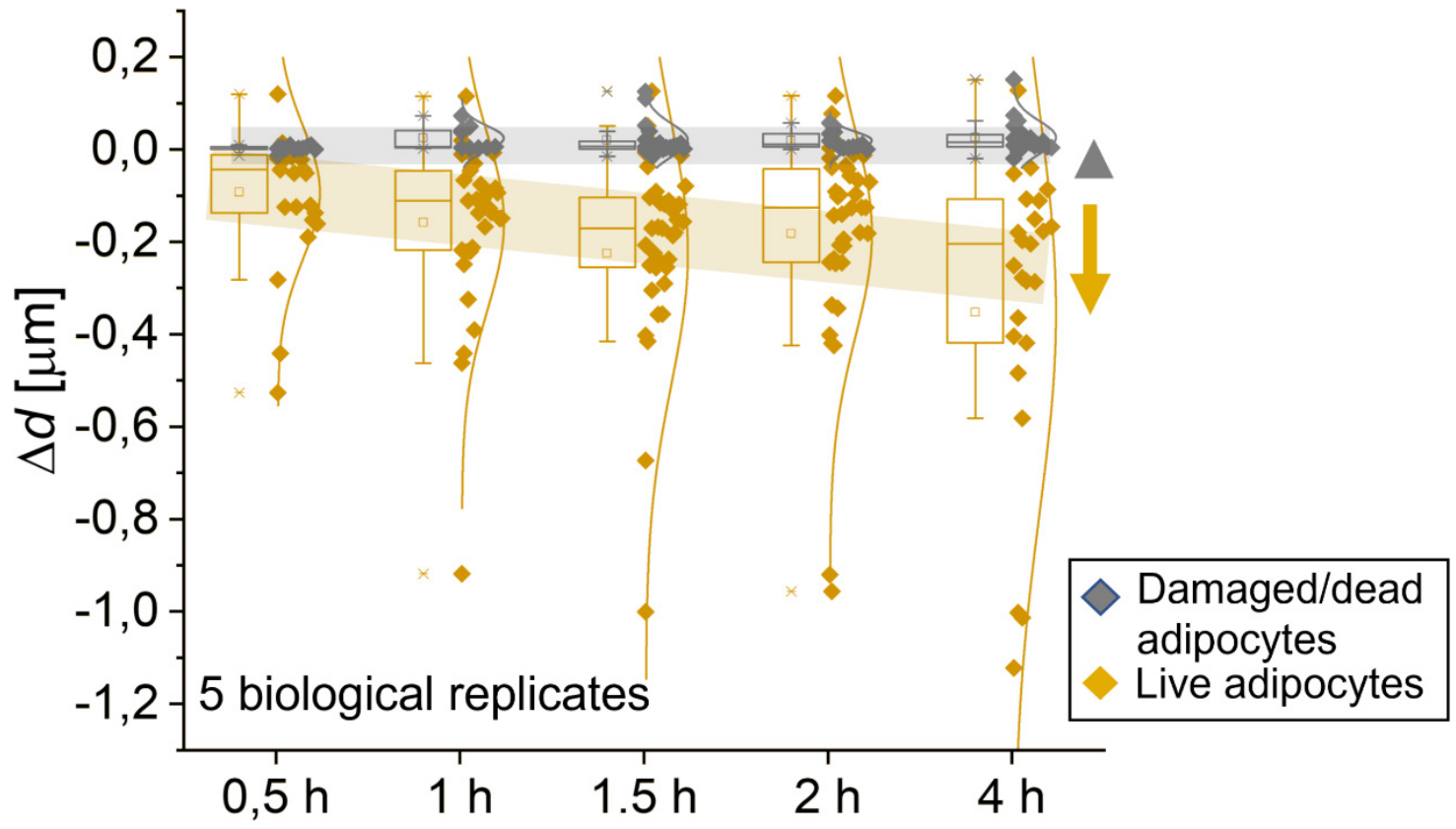

**Figure 5.** A comparison and a clear difference in the dynamics of LDs size change (Δ*d*) between live and damaged adipocytes obtained from 5 biological replicates, each analyzing several cells. The sample size for each biological replicate was between 3 and 8.

## Conclusions and future outlook

We present a novel experimental methodology for precisely sensing adipocyte size dynamics with exceptionally high spatial (sub-nanometer) and temporal (sub-minute) resolution. This approach builds on our earlier study, where we demonstrated lasing inside live cells for the first time [14]. We systematically investigated the physical and biological aspects of photonic sensing in live, mature adipocytes, highlighting both its capabilities and limitations. Due to the non-spherical and uneven morphology of LDs in live adipocytes, conventional methods for generating WGM resonances proved inadequate. Instead of using a pulsed nanosecond laser—which requires high lasing powers and risks photo-induced damage—we implemented a cost-effective, low-power CW laser. Although this setup did not yield the maximum Q-factor or spectral resolution typical of pulsed lasers, it was not a limiting factor. The temporal spectral noise was comparable to the peak position uncertainty from spectral fitting, resulting in a precision of approximately 3.5 nm—two orders of magnitude better than the optical resolution achievable with confocal microscopy. Using precise WGM spectral characterization and appropriate fitting models, we quantified biological and morphological heterogeneity in adipocytes and captured rapid metabolic responses to external lipolytic stimuli. These responses were measurable through changes in LD size and molecular flux rates. Moreover, WGM spectral shifts proved to be a promising tool for rapid diagnostics of cellular state and viability. Our results establish a proof-of-concept that significantly advances current lipolytic and viability assays by enabling faster, more cost-effective and single-cell–level assessment of primary adipocyte dynamics and heterogeneity, which are indispensable for understanding genuine metabolic responses. Unlike conventional assays, which rely on bulk cell populations and cannot resolve interactions between individual cells, our methodology enables such analysis. If adapted for high-throughput imaging and further combined with emerging single-cell RNA-sequencing techniques [32,60,71], this complementary approach holds broad potential for investigating the mechanisms of metabolic and obesity-related diseases at both cellular and tissue scales. It would substantially increase sample size and statistical power, improving precision in characterizing biological and morphological heterogeneity and enabling more robust insights into fast transient and early kinetic-dynamic processes within adipose tissue.

## Data Availability

The raw data underlying this study are openly available in the Zenodo repository at DOI: 10.5281/zenodo.17806820.

## Supporting Information

Supporting Information is available free of charge at:

- Fig S1: WGM spectra preparation and analysis to calculate change in LD size; Fig S2: examples of spectral fits to differently-shaped WGM spectra; Fig S3: validation of LD size change with optical measurements; Fig S4: diffusive transport of lipolytic agents through a trans-well system used in our study; Fig S5: examples of adipocyte preservation and viability after isolation from mice visceral fat; Fig S6: evaluation of the viability of adipocytes; Fig S7: performance of pulsed and CW pump laser sources for safe WGM-based biosensing applications on adipocytes; Fig S8: measurement of the LD sphericity; Fig S9: measurement of the spectral repeatability on a stable LD; Fig S10: transient effect on individual LD in the rate of size change ($\Delta d/t$) after isoproterenol stimulation; Fig S11: comparison of WGM spectra-based rapid diagnostics of adipocyte viability using pulsed and CW laser sources; Fig S12: analysis and fitting of WGM spectra in a perturbed adipocyte with a significant spectral shape change: Fig S13: an example of significantly increased TE and TM mode splitting in the WGM spectra due to plasma membrane rupture;
- Supp. Note A: comment on plasma membrane remodeling; Supp. Note B: detailed description of WGM peak uncertainty calculation.

## Acknowledgement

This work was supported by the European Research Council (ERC) under the European Union's Horizon 2020 research and innovation program (ERC Starting Grant, GA851143) and by the Slovenian Research and Innovation agency (ARIS) (P1-0099, N1-0362 and P1-0140).

## Authors Contributions

R.P. conducted and designed the experiments, analyzed the results and wrote the manuscript with input from other authors; A.K. prepared the designed the sample preparation; M.Z. conducted initial experiments; P.M.Š. prepared the samples; S.U. supported original idea and supervised the study; M.H. conceived the original idea, designed, and supervised the study.

## Supplementary Data

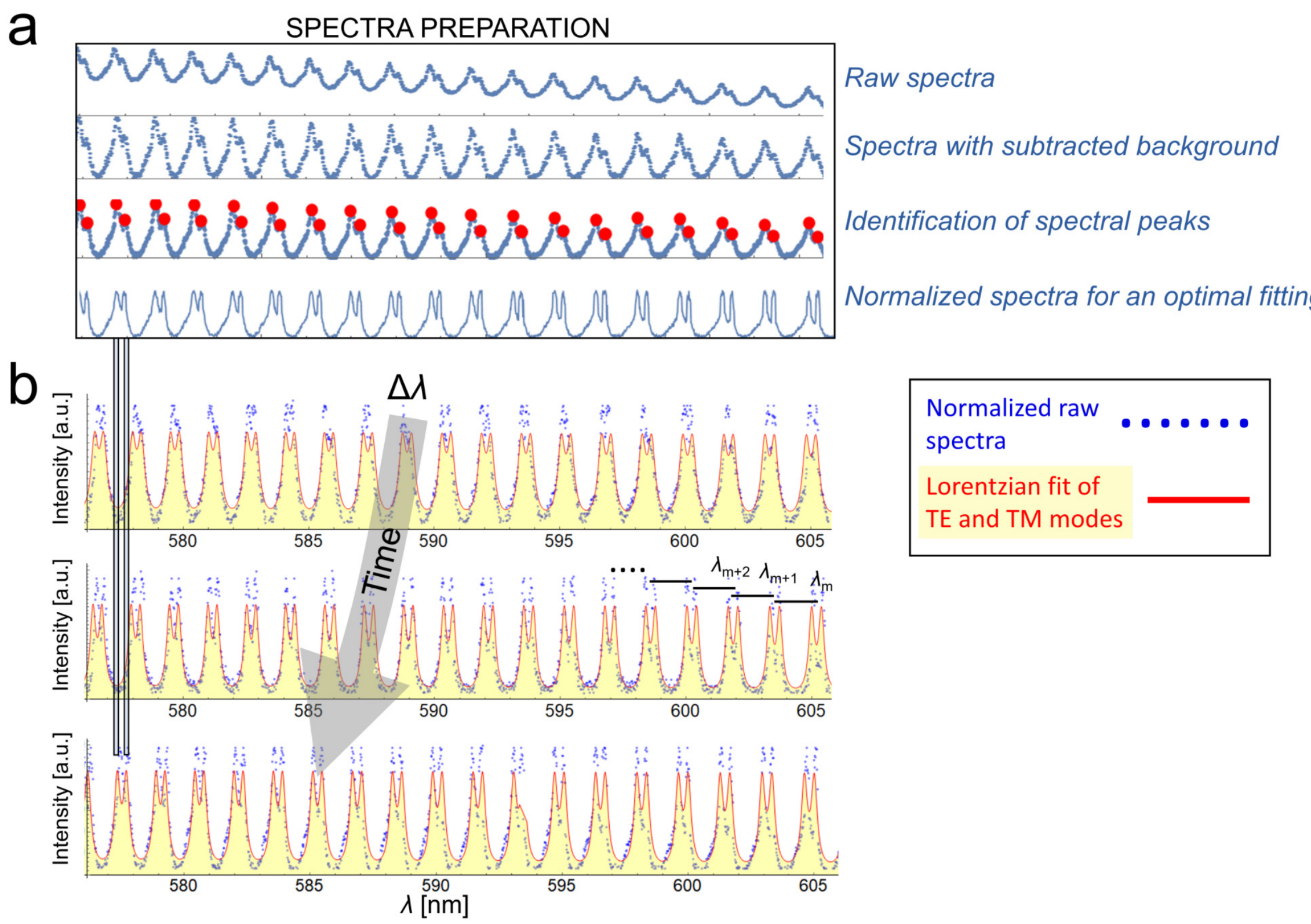

**Figure S1.** Consecutive steps of WGM spectra preparation and analysis to calculate change in LD size. (a) Spectra preparation for the optimized fitting. (b) Fitting of normalized spectra by using first order radial mode approximation description and Lorentzian function for TE and TM eigenmodes (equation 1) in the time experiment.

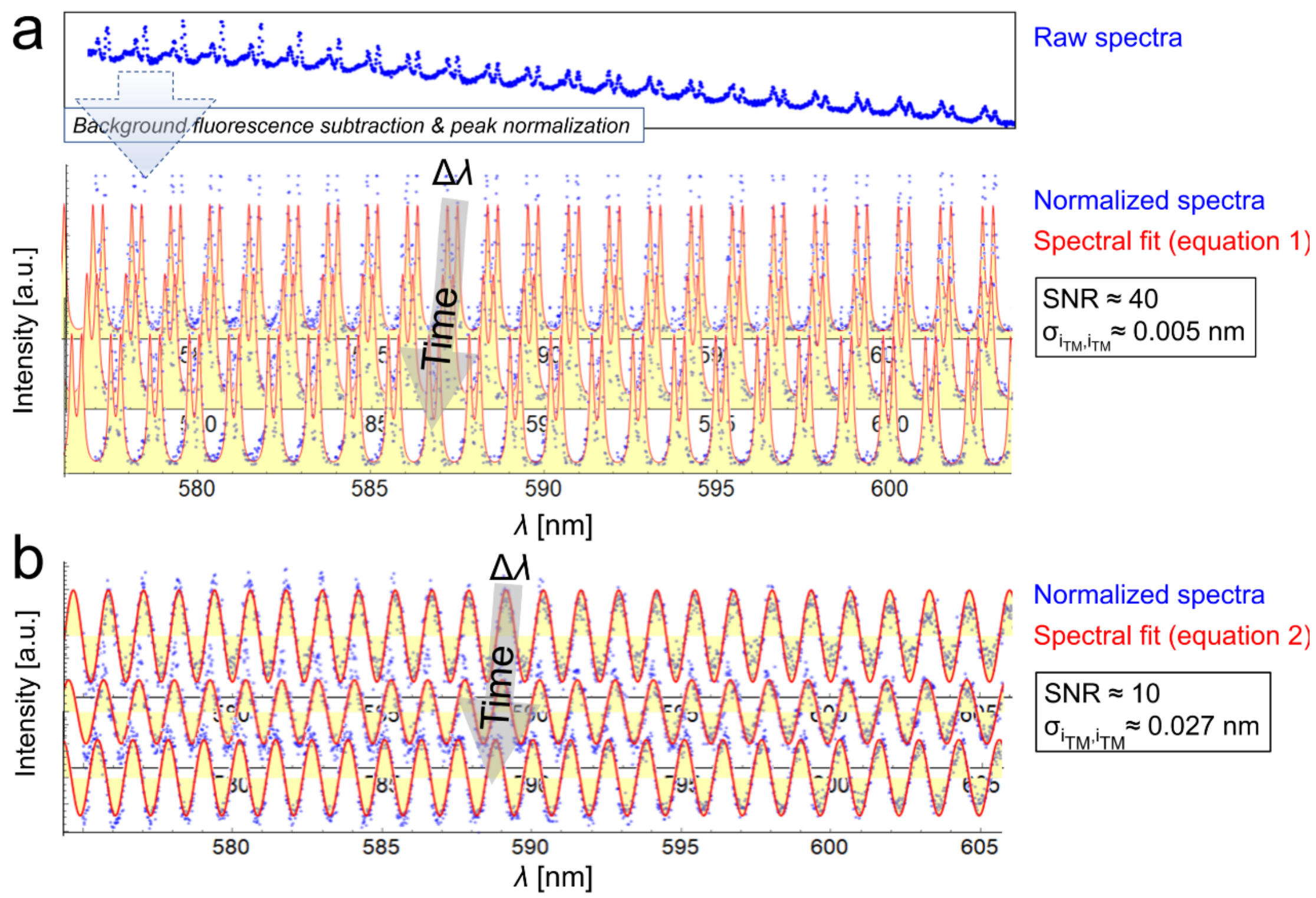

**Figure S2.** An example of different spectral fits to differently-shaped, background-subtracted and normalized raw WGM spectral data in a time experiment. (a) The workflow for precise spectral fitting and analysis of distinctly observed TM and TE eigenmodes providing the spectral peak uncertainty of $\sigma_{iTM,iTM} \approx 0.005$ nm for ultraprecise characterization of adipocyte size. (b) Precise spectral fitting and analysis of the WGM spectra with an extensive spectral broadening providing the spectral peak uncertainty of $\sigma_{iTM,iTM} \approx 0.027$ nm.

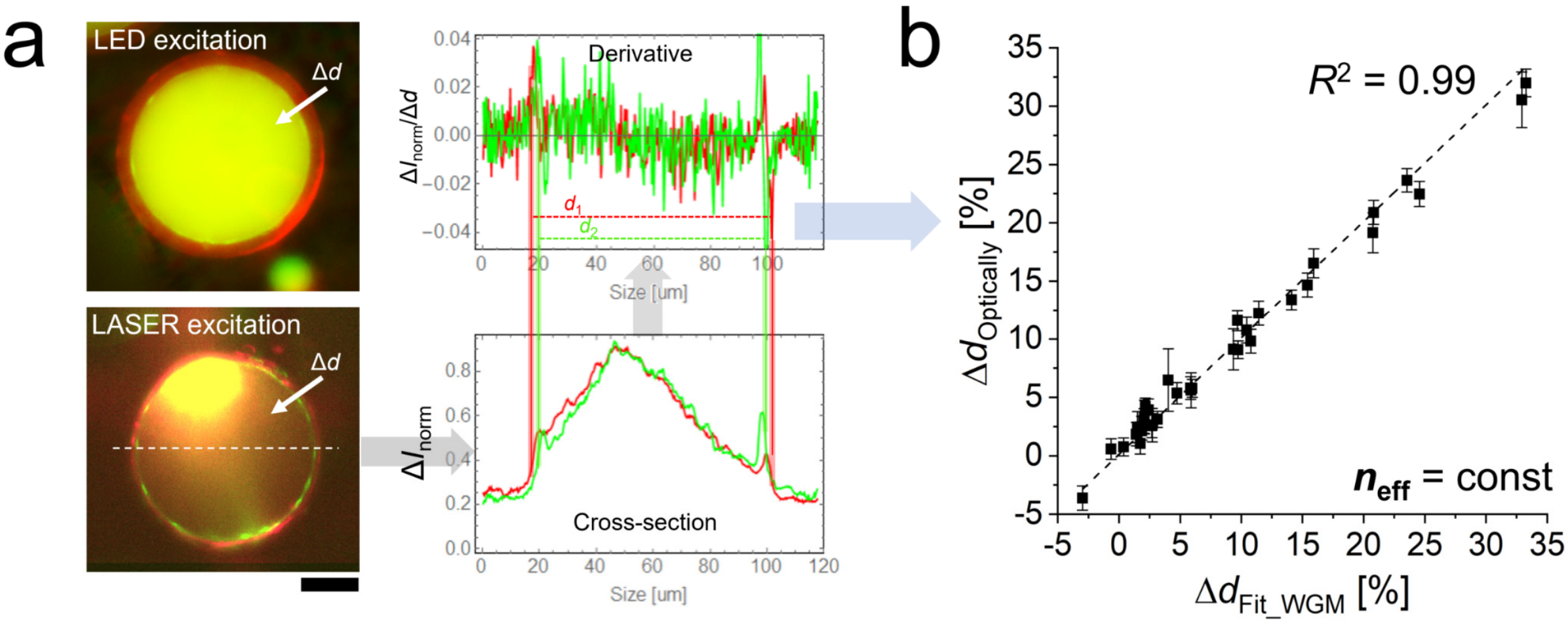

**Figure S3.** Optical measurements of LD size change ($\Delta d$) feasible in adipocytes with prolonged lipolytic activity. (a) An example of a significant decrease in LD size after one day, measured through the fluorescence signal of Pyromethene 597, artificially color-coded in red

and green (left), and by image analysis of the size change obtained from the derivative of the intensity cross-section profile. (b) Comparison of optical measurements of $\Delta d$ (in %) with values calculated from WGM fitting, under the assumption of a constant $n_{eff}$ over time, revealed excellent correlation across the entire measured range ($R^2$ = 0.99).

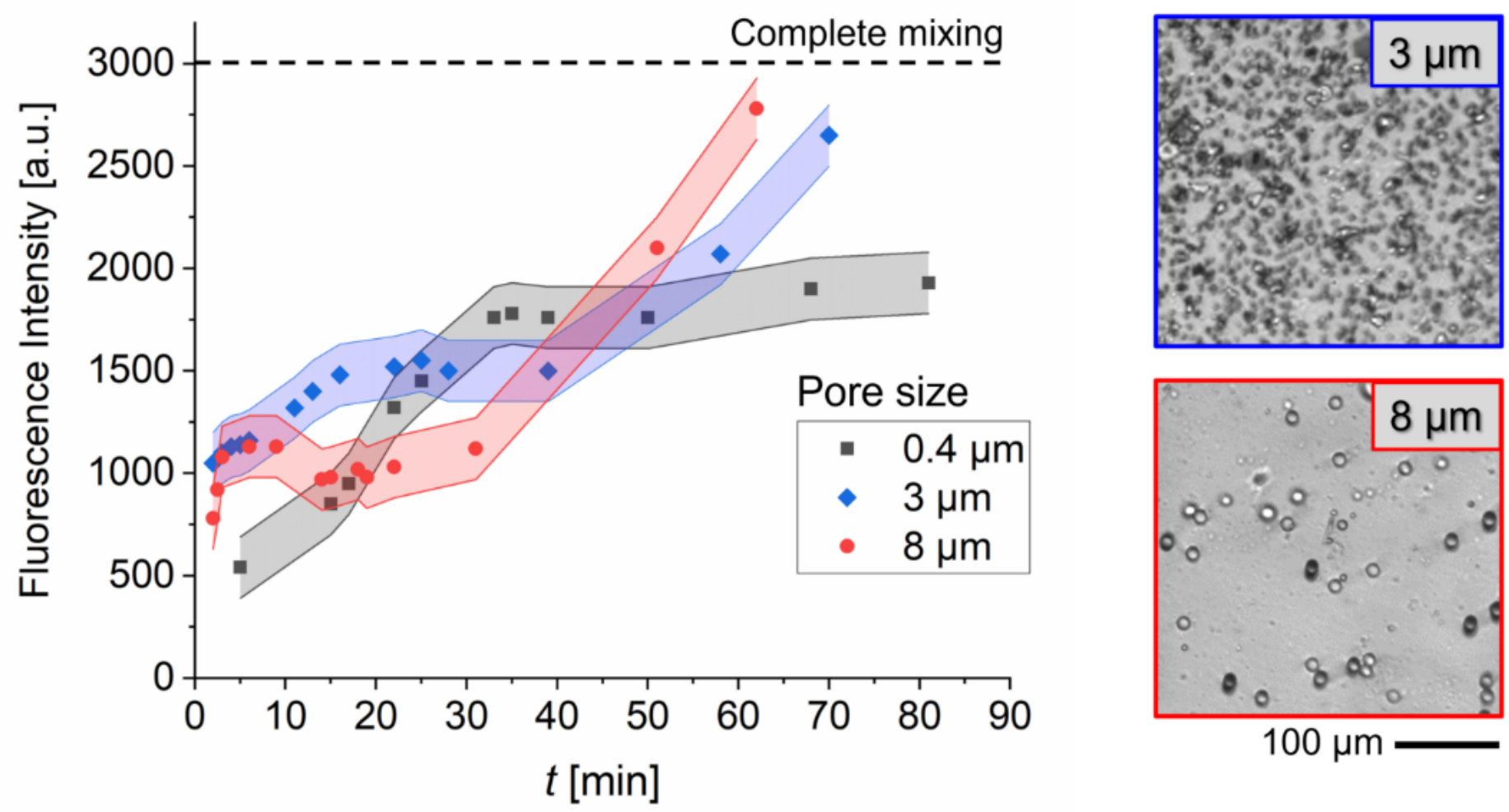

**Figure S4.** Diffusive transport of small molecules through different-sized pores in a trans-well system to reach complete mixing between both chambers. Faster transport of molecules through smaller 3 µm pores as compared to 8 µm pores in the first 30 min is attributed to the much higher density of smaller pores, as seen in the right images. The Trans-well with the smallest pores (0.4 µm in gray) did not achieve complete mixing within 90 minutes and was thus not used in our study. On the other side, both 3 µm and 8 µm trans-well chambers were appropriate for our multi-hour dynamics study on adipocytes. The color-coded bands show the error of the measurements.

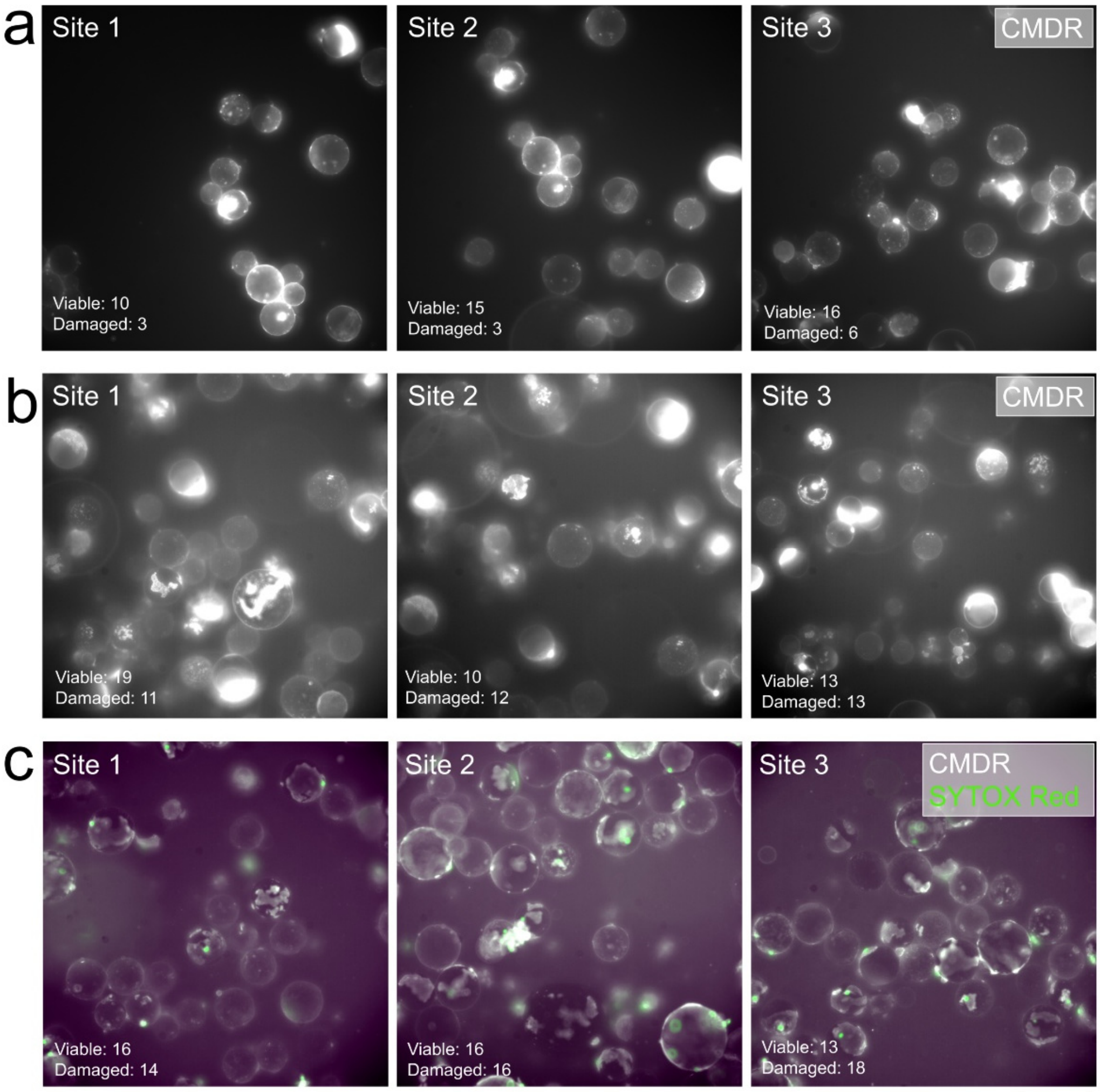

**Figure S5.** Adipocyte preservation and viability measurements after being isolated from visceral AT presented on different biological replicates (a-c). a) An estimated viability of 70-80% using CellMask Deep Red plasma membrane stain to track membrane rupture and localized accumulation of the stain. b) An estimated viability of 50-70% using the same stain. c) An estimated viability of 40-60% using the combination of CellMask Deep Red (in gray) and SYTOX Red nucleic acid stain (in green), with the distinct features nicely colocalized on the damaged structures. Scale bar is 100 μm.

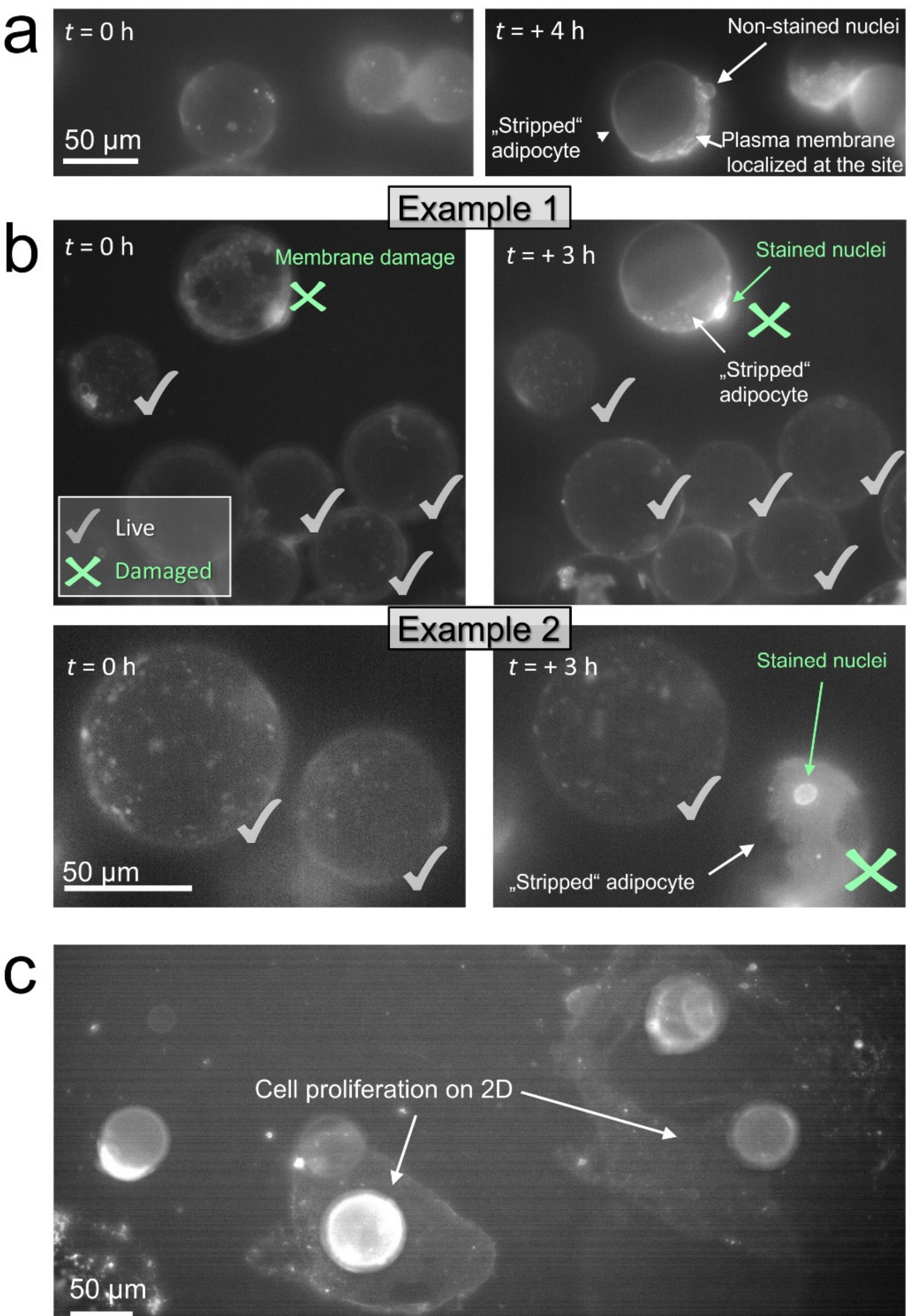

**Figure S6.** Evaluation of the viability of isolated mature adipocytes for further LD dynamics experiments. (a) Ruptured plasma membrane with possibly retained structural integrity due to non-stained nuclei. (b) The two exemplary cases where the remodeled/ruptured plasma membrane correlates with nuclear labeling, indicating cellular damage/death (marked in green). Adipocytes with a homogenously stained plasma membranes (checkmarks) remain highly viable, as indicated by the live/dead assay. (c) Cell proliferation on the transwell surface indicates adipocyte dedifferentiation. Imaging was performed using a 20× objective (*NA* = 0.45) with fluorescent labeling of the plasma membrane (CellMask Deep Red, Invitrogen) and nuclei (SYTOX Deep Red).

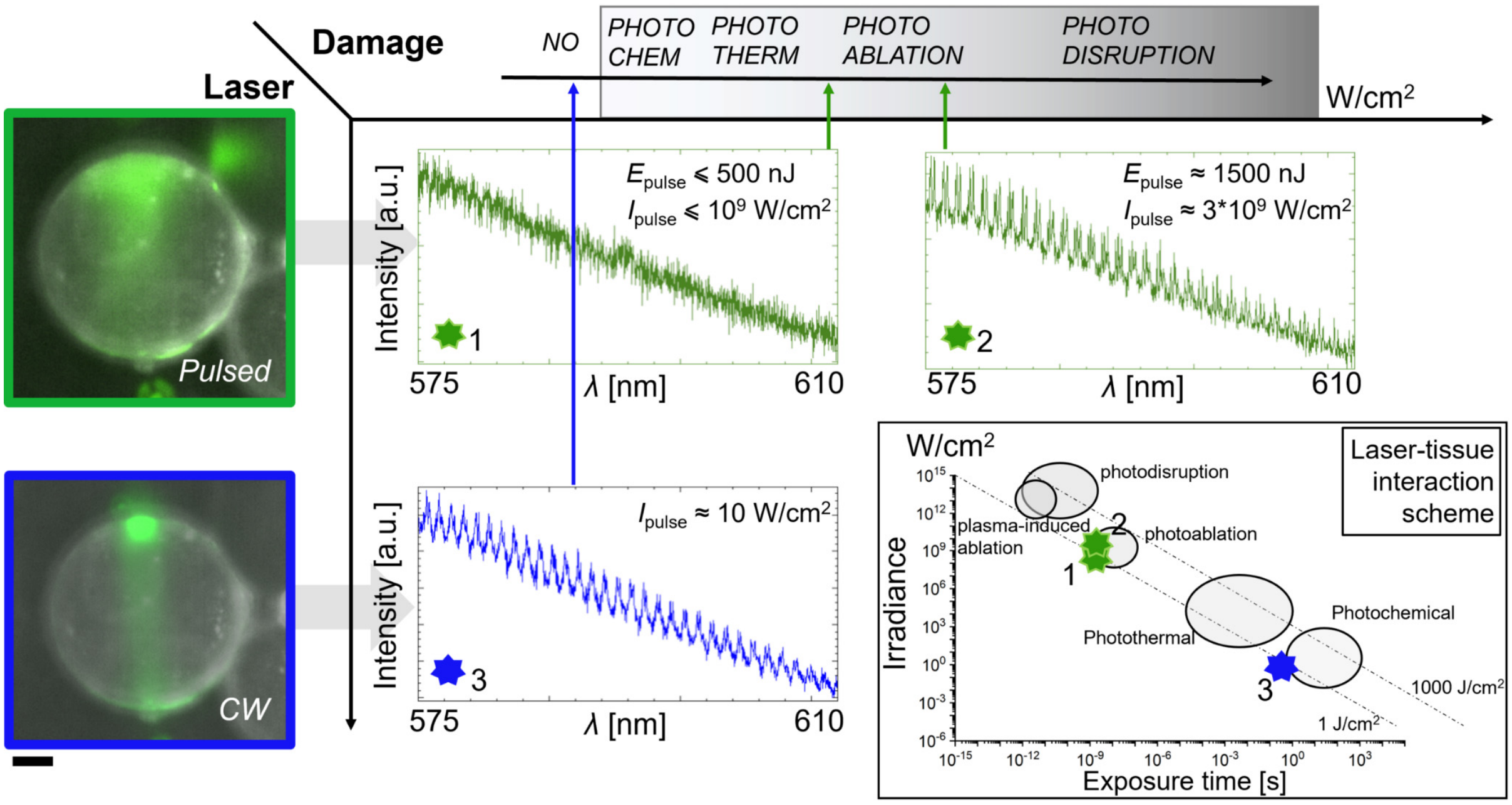

**Figure S7.** Performance of pulsed (ns) and CW pump laser sources for safe WGM-based biosensing applications on mature adipocytes. Due to the non-ideal sphericity and smoothness of LDs in live adipocytes, which are locally disturbed by the cytoskeleton and the imbalance of surrounding mechanical forces, high laser doses/irradiance per pulse ($E_{pulse}/I_{pulse}$), capable of inducing a photoablation-damaging effect (green arrow), are commonly required to introduce WGM lasing on adipocytes (2 - green spectra on the right). By introducing an alternative approach using a cost-efficient CW laser, slightly lower performance in spectral resolution and SNR was achieved (3 - blue spectra), but with negligible damaging effects, as shown with the blue arrow and schematically depicted in the laser-tissue interaction map on the bottom right (adapted from [1]). Scale bar is 10 μm.

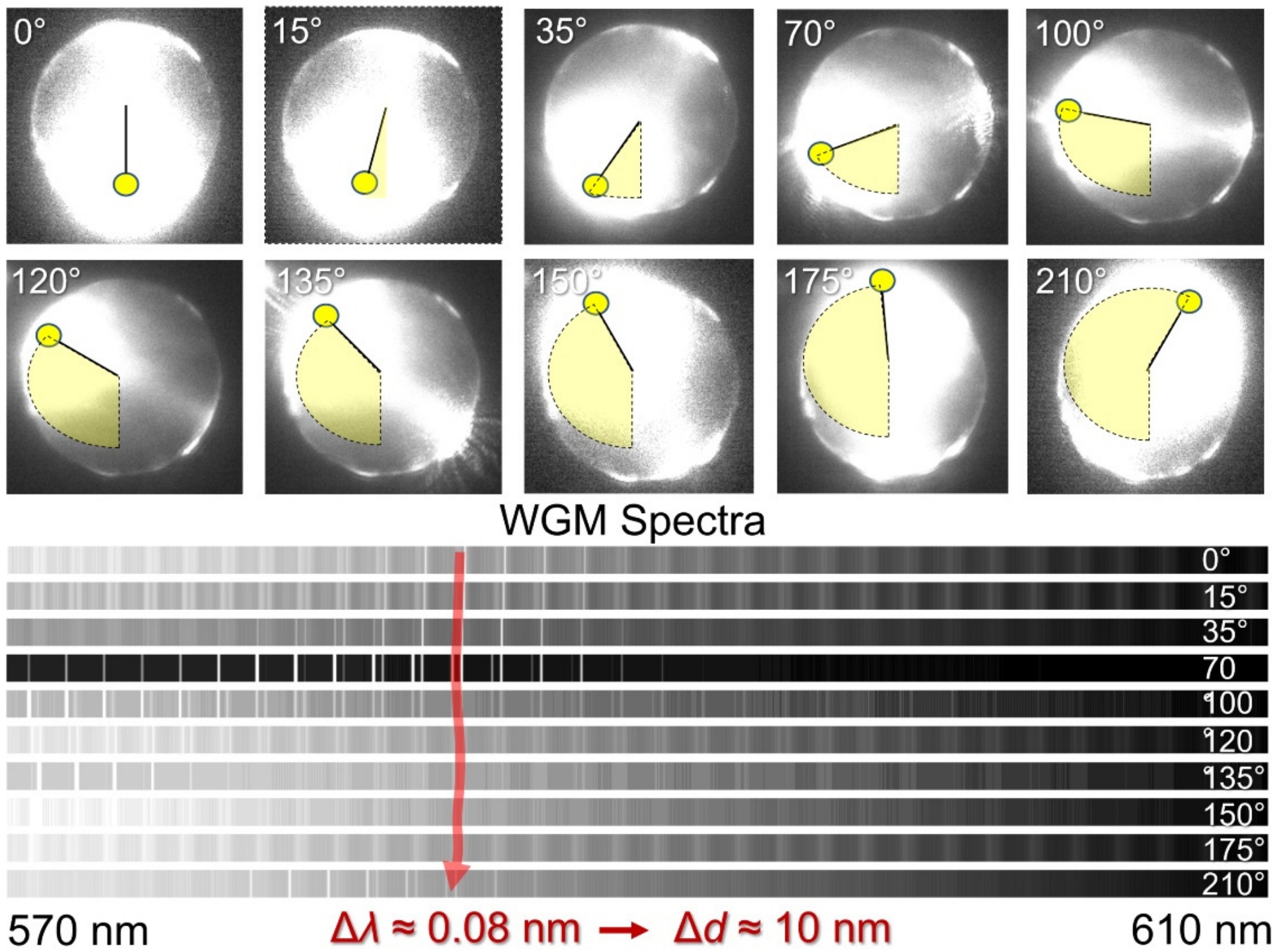

**Figure S8.** Measurement of the LD sphericity by changing the pump laser position around its circumference. Acquired spectra at different positions, from angle 0° to 210°, reveal deviation from complete sphericity for up to $\Delta\lambda$ = 0.08 nm (red arrow), which accounts for the diameter change of up to $\Delta d$ = 10 nm, lower than typical changes induced by external stimuli.

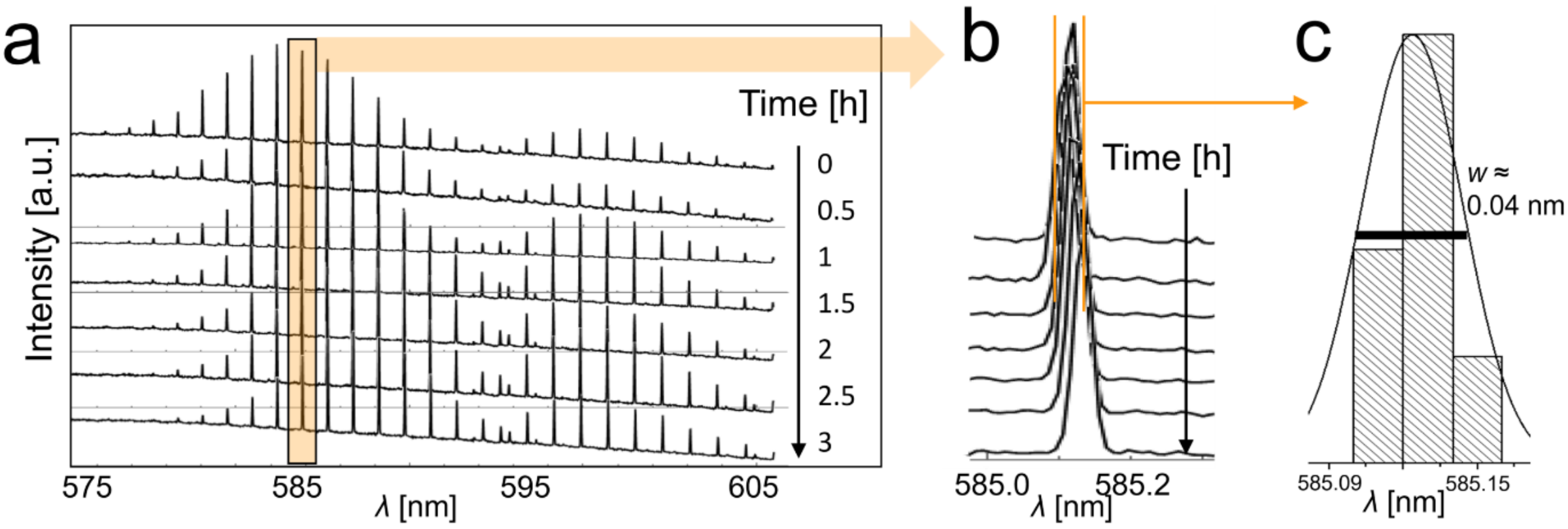

**Figure S9.** Measurement of the scatter of the multiple spectra acquired from a stable LD throughout a typical time experiment using a pulsed laser. We used a pulsed laser because it provides WGMs with a higher Q-factor than is achievable with a CW laser, thereby enabling greater measurement precision. (a) WGM spectra performed in a 3-hour time interval with 0.5-hour time steps. b) A slight scatter and a directional trend (red shift) of the spectra as shown on a single eigenmode at $\lambda$ = 585.13 nm. c) Distribution of peak positions gathered from multiple spectra.

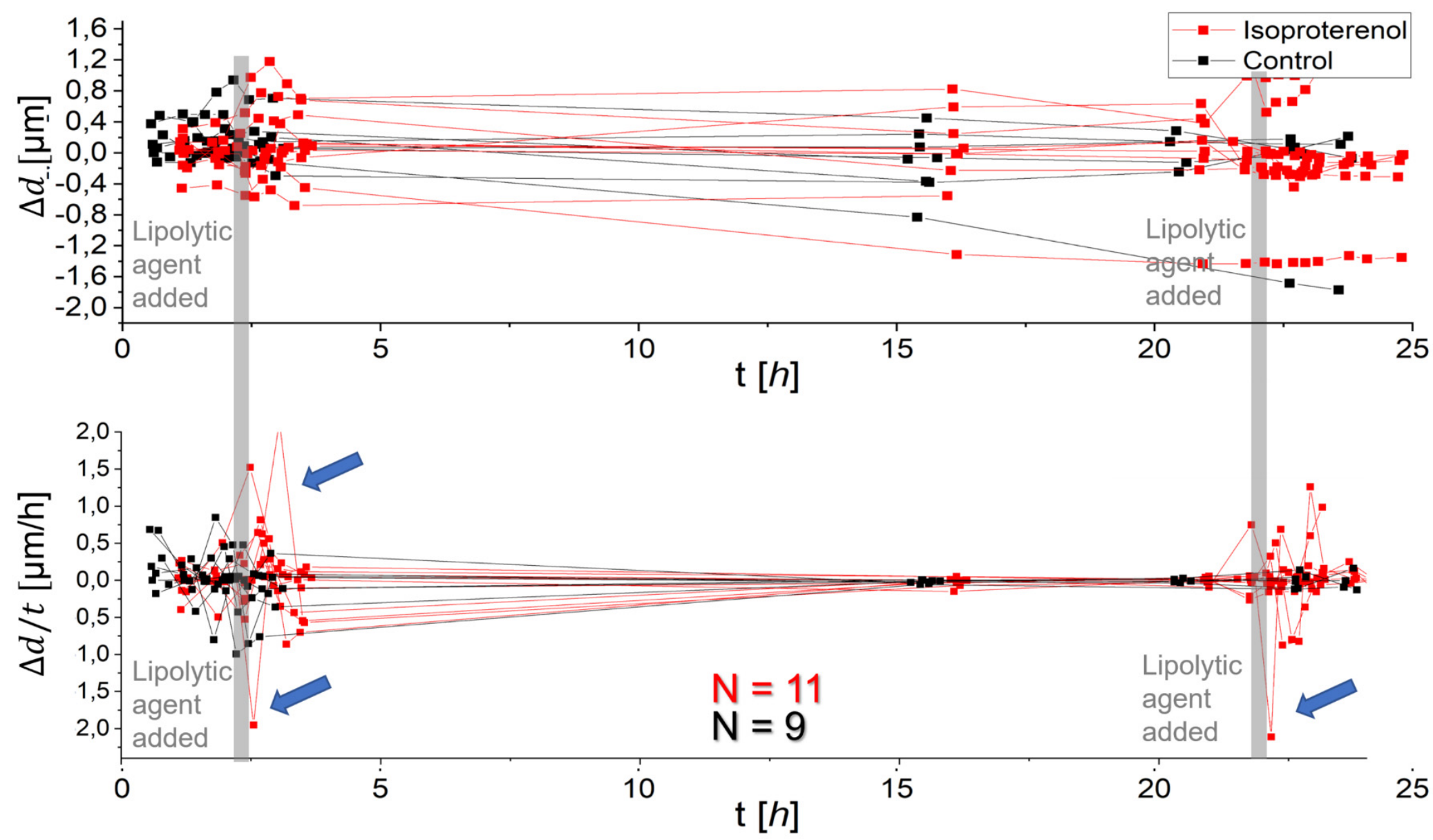

**Figure S10.** Adipocyte metabolic response via measurements of LD size ($\Delta d$) and rate of size change ($\Delta d/t$) to external stimuli induced by isoproterenol (connected data points in red). The data reveal adipocyte heterogeneity and a rapid, transient effect on individual adipocytes (marked with the blue arrow), where the rate of $\Delta d$ oscillates significantly more than in control.

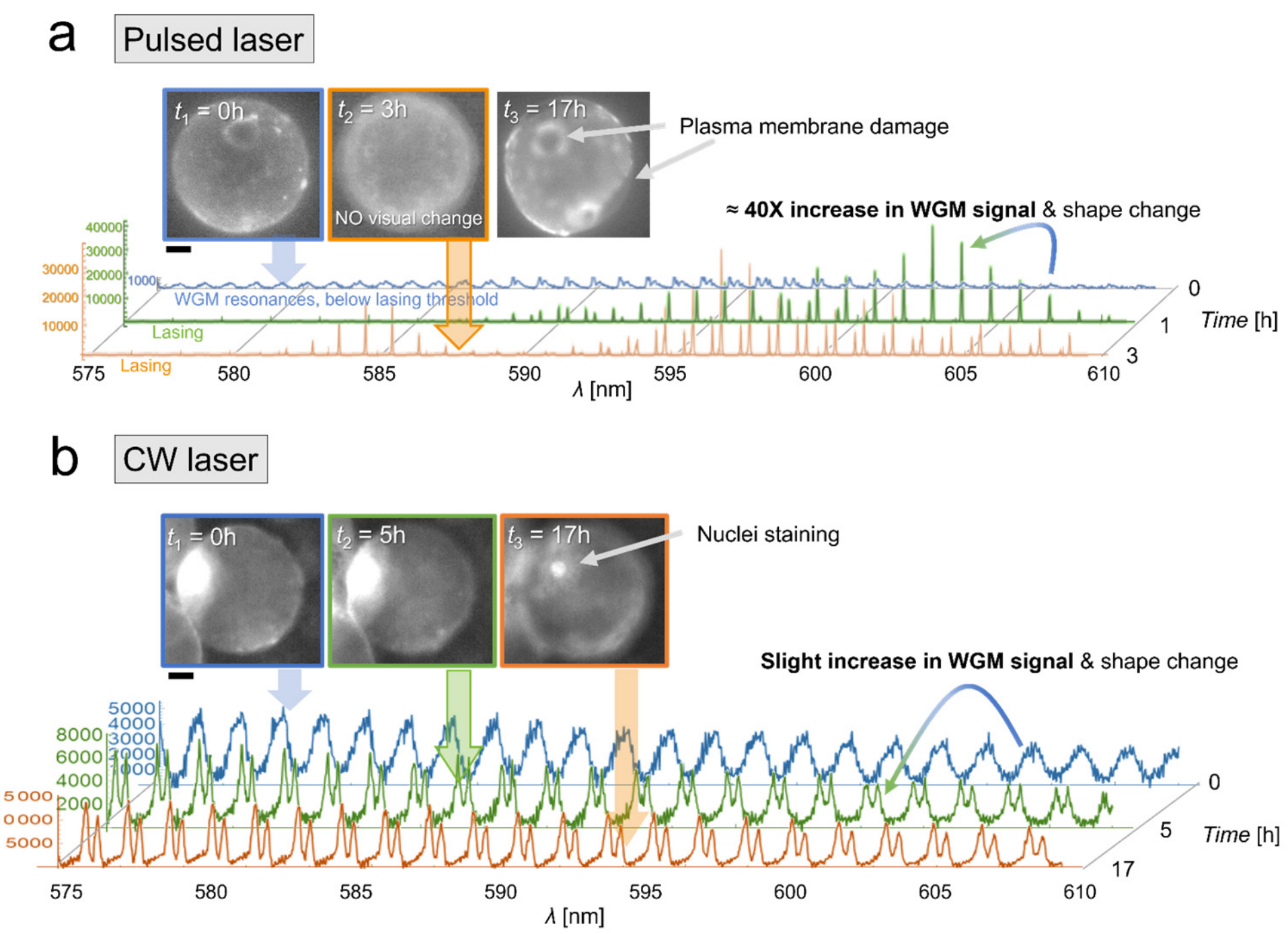

**Figure S11.** Comparison of WGM spectra-based rapid diagnostics of adipocyte viability using pulsed and CW laser sources, demonstrating superior spectral sensitivity with the pulsed laser. (a) Fluorescence images of the same cell at three time points, stained with the plasma membrane dye CellMask Deep Red, together with the corresponding WGM spectra induced by pulsed laser excitation, color-coded in blue, green, and orange. (b) Fluorescence images of the same cell at three time points, stained with CellMask Deep Red in combination with the nuclear viability dye SYTOX Deep Red, together with the corresponding WGM spectra induced by CW laser excitation, color-coded in blue, green, and orange. Scale bar: 10 µm.

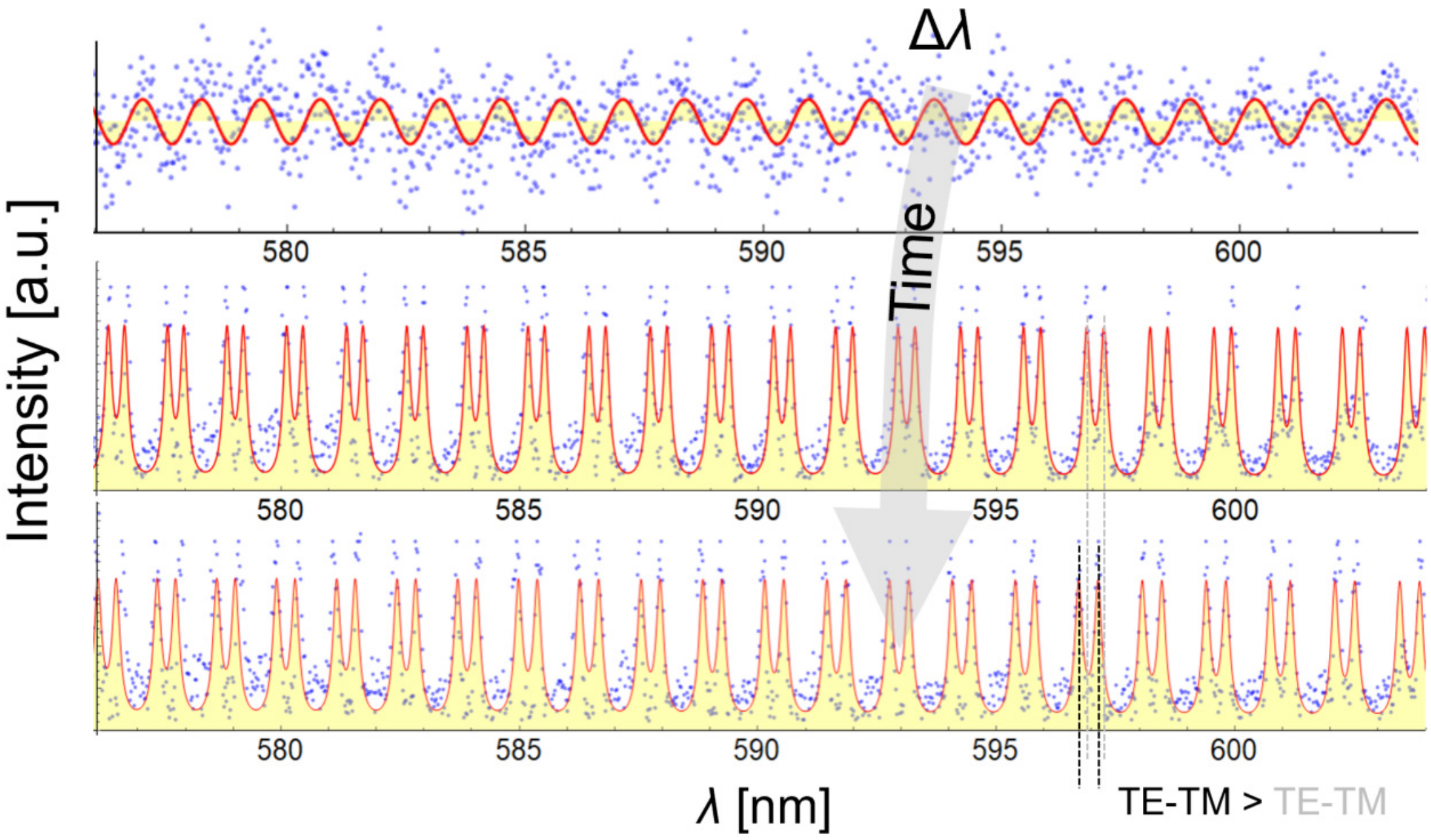

**Figure S12.** Detailed analysis and fitting of WGM spectra (in red) with an empirical (upper panel; equations 3 and 4) and exact (lower panels; equation 1) model for the measurement of LD size in a perturbed adipocyte. A significant spectral shape change was observed between the first and second time points, indicating cellular damage or plasma membrane rupture, which was later confirmed optically (see Figure 5c). A detailed analysis has also revealed a slight but measurable increase in TE and TM mode splitting (black and gray dashed lines), characterized by a decrease in the refractive index in the cellular cytoplasm, indicating dilution induced by plasma membrane rupture.

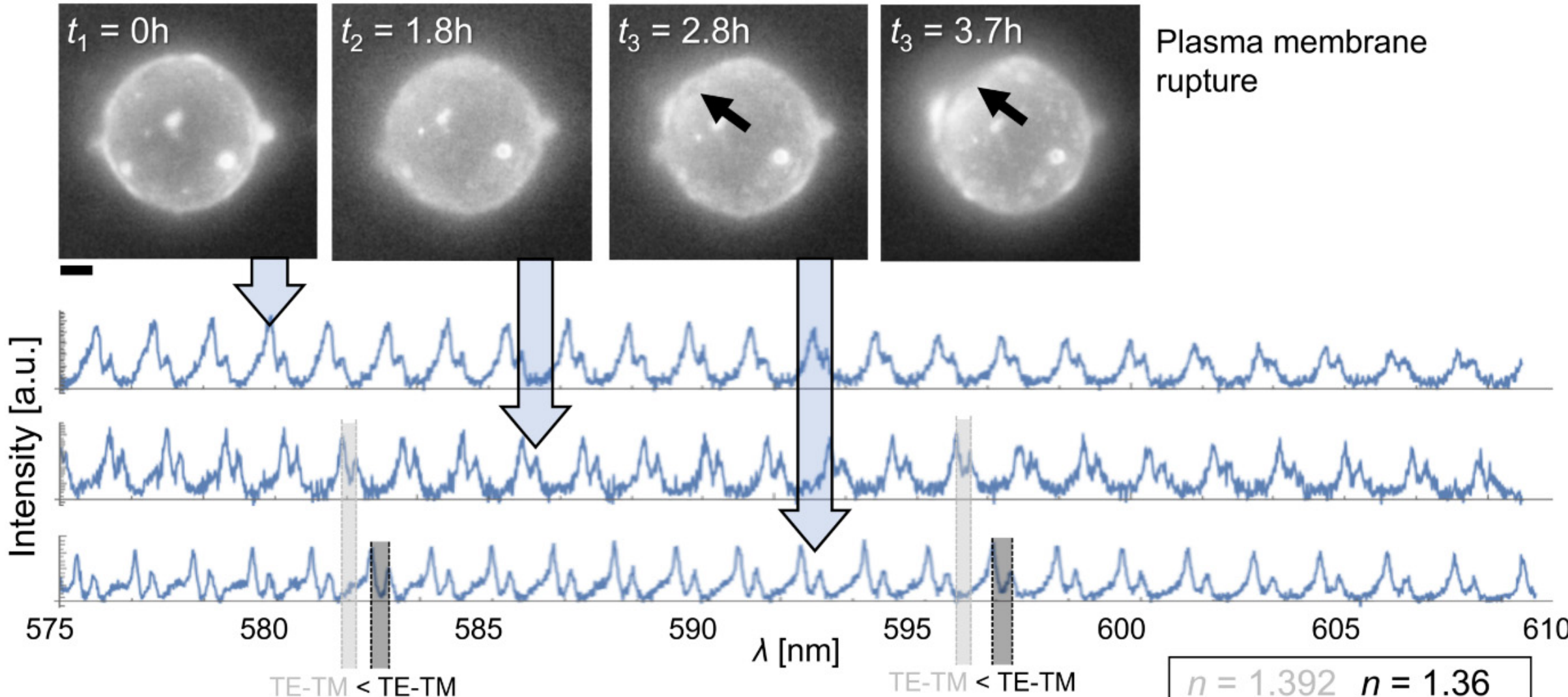

**Figure S13.** An example of significantly increased TE and TM mode splitting in the WGM spectra is shown between the second and third time points (see the width of the gray bars showing the splitting). This spectral change, characterized by a measurable decrease in the refractive index of the cellular cytoplasm around the LD, is correlated with plasma membrane rupture, as nicely observed in the last time point (see the black arrow). Scale bar is 10 µm.

## Supplementary Note A

In certain cases of cellular disruption, where the plasma membrane appeared localized to one side of the “stripped” adipocyte, the membrane seemed to retain structural integrity. This was evident from the presence of unstained nuclei (Figure S5a), which under normal conditions would be stained (Figure S5b). Such observations suggest a potential, albeit unconventional, mechanism for preserving the function of core cellular components under mechanical, physical, or metabolic stress. In contrast, lipid droplets (LDs), once partially secreted, exhibited no dynamic activity, indicating arrested metabolism and/or cellular damage, as discussed later. Plasma membrane remodeling may therefore reflect not only cellular damage but also contribute to adipocyte dedifferentiation [2], as indicated by cell proliferation in 2D culture (Figure S5c). Given the complexity of the biological system under study, we focused exclusively on mature, non-differentiated adipocytes that maintained both plasma membrane and LD integrity. In this context, staining of both nuclei and plasma membrane was essential.

## Supplementary Note B

By applying spectral fitting to eigenmodes with profiles resembling Gaussian shapes, and using a well-defined model for peak position error (dependent on spectrometer resolution, sampling density, and SNR) [3] we calculated the uncertainty ($\sigma_{\lambda_{i_{TM},i_{TE}}}$) [4], and thereby the resolution in peak position. This uncertainty is expressed by the following equation:

$$\sigma_{\lambda_{i_{TM},i_{TE}}} = \frac{\sqrt{\Delta\lambda_s W}}{\text{SNR}}\sqrt{\frac{\Delta}{F(\Lambda)}},$$

where $\boldsymbol{\Delta\lambda_s}$ is the wavelength sampling step ($\boldsymbol{\Delta\lambda_s}$ = 0.02 nm); *W* is related to the spectral full width at half maximum (FWHM; *w*) as *W* ≈ *w*/2.35; SNR is defined as the maximum signal value at each eigenmode divided by the spectral noise; *Δ* is a numerical constant representing the confidence limit for a one-parameter fit (*Δ* = 1 corresponding to a 68% confidence interval or "1σ" error); *F*(Λ) is a dimensionless factor derived from the integral of $(\boldsymbol{\partial S}_{\text{eigenmode}}/\boldsymbol{\partial\lambda_{i_{TM},i_{TE}}})^2$ over the *λ*-acquisition range (*Λ*), where $\boldsymbol{S}_{\text{eigenmode}}$ denotes the spectral profile of the individual eigenmode. Given that our acquisition range spans the full spectrum, $\boldsymbol{F^{1/2}}$ assumes a value of approximately 0.8 under the assumption of Gaussian noise [3]. Prior to calculating $\boldsymbol{\sigma_{\lambda_{i_{TM},i_{TE}}}}$, the equation was slightly adjusted to account for oversampling in $\boldsymbol{\Delta\lambda_s}$ relative to the spectrometer resolution (0.06 - 0.07 nm). This oversampling effectively reduced the number of independent data points across the acquisition range, leading to a modest increase in peak position uncertainty, as supported by analytical and numerical evaluations in study [3]. For the spectra shown in Figure 2g and Figure S2a—with an SNR of approximately 40 and *w* ≈ 0.15 nm—the WGM peak uncertainty was calculated to be $\boldsymbol{\sigma_{\lambda_{i_{TM},i_{TE}}}}$ ≈ 0.005 nm.